\documentclass[fleqn,usenatbib]{rasti}
\hypersetup{nolinks=true}

\usepackage{newtxtext,newtxmath}

\usepackage[T1]{fontenc}

\DeclareRobustCommand{\VAN}[3]{#2}
\let\VANthebibliography\thebibliography
\def\thebibliography{\DeclareRobustCommand{\VAN}[3]{##3}\VANthebibliography}

\usepackage{graphicx}	
\usepackage{amsmath}	
\usepackage{blkarray}

\title[Entropy of galaxy spectra]
{The entropy of galaxy spectra: How much information is encoded?}

\author[Ferreras et al.]
{Ignacio Ferreras$^{1,2,3}$\thanks{E-mail: i.ferreras@ucl.ac.uk},
Ofer Lahav$^{1}$, Rachel S. Somerville$^{4,5}$, Joseph Silk$^{6,7,8}$\\
$^1$ Department of Physics and Astronomy, University College London, London WC1E 6BT, UK\\
$^2$ Instituto de Astrof{\'i}sica de Canarias, Calle V{\'i}a L{\'a}ctea s/n,
E38205, La Laguna, Tenerife, Spain\\
$^3$ Departamento de Astrof{\'i}sica, Universidad de La Laguna, E38206 La Laguna, Tenerife, Spain\\
$^4$ Center for Computational Astrophysics, Flatiron Institute, New York, NY 10010, USA\\
$^5$ Department of Physics and Astronomy, Rutgers University, 136 Frelinghuysen Rd, Piscataway, NY 08854, USA\\
$^6$ Institut d'Astrophysique de Paris, UMR7095:CNRS \& UPMC, Sorbonne University, F-75014, Paris, France\\
$^7$ Department of Physics and Astronomy, The Johns Hopkins University Homewood Campus, Baltimore, MD 21218, USA\\
$^8$ Beecroft Institute of Particle Astrophysics and Cosmology, Department of Physics, University of Oxford, Keble Ro
ad, Oxford OX1 3RH, UK
}

\date{RASTI: Accepted 20/01/2023 . Received 11/11/2022 ; in original form 27/05/2022}

\pubyear{2023}

\begin{document}
\label{firstpage}
\pagerange{\pageref{firstpage}--\pageref{lastpage}}
\maketitle

\begin{abstract}
  The inverse problem of extracting the stellar population content of
  galaxy spectra is analysed here from a basic standpoint based on
  information theory.  By interpreting spectra as probability
  distribution functions, we find that galaxy spectra have high
  entropy, thus leading to a rather low effective information content.
  The highest variation in entropy is unsurprisingly found in regions
  that have been well studied for decades with the conventional
  approach. We target a set of six spectral regions that show the
  highest variation in entropy -- the 4,000\AA\ break being the most
  informative one. As a test case with real data, we measure the
  entropy of a set of high quality spectra from the Sloan Digital Sky
  Survey, and contrast entropy-based results with the traditional
  method based on line strengths. The data are classified into
  star-forming (SF), quiescent (Q) and AGN galaxies, and show --
  independently of any physical model -- that AGN spectra can be
  interpreted as a transition between SF and Q galaxies, with SF
  galaxies featuring a more diverse variation in entropy.  The high
  level of entanglement complicates the determination of population
  parameters in a robust, unbiased way, and affect traditional methods
  that compare models with observations, as well as machine learning
  (especially deep learning) algorithms that rely on the statistical
  properties of the data to assess the variations among spectra.
  Entropy provides a new avenue to improve population synthesis models
  so that they give a more faithful representation of real galaxy
  spectra.
\end{abstract}

\begin{keywords}
methods: data methods --
methods: statistical --
techniques: spectroscopic --
galaxies: stellar content
\end{keywords}

\section{Introduction} \label{sec:intro}
\label{Sec:Intro}

Stars and gas constitute the most fundamental observables in
extragalactic astrophysics.  While the gaseous component, in its
different guises, mostly reflects the ongoing physical state, the
stellar populations encode valuable information about the past
formation history, both through their collisionless nature -- that
keep track of the past dynamical history -- and their age and chemical
composition -- that trace the past star formation history. The most
relevant spectral window to probe the stellar content is the rest-frame
NUV/optical/NIR range, where most of the light from stellar photospheres
is emitted. The standard way to study the underlying
stellar population focuses on comparisons of photo-spectroscopic
observables with population synthesis models
\citep[to name a few,][]{BC93,Wo:94,BC03,Pegase,Maraston:05,Vazdekis:10,FSPS:10}, 
that combine our understanding of stellar formation and evolution, along
with a determination (empirical, theoretical or mixed) of the stellar
atmospheres \citep[see, e.g.][for a general view of these
  models]{Walcher:11,Conroy:13}.

While this paper focuses on an alternative
  analysis of galaxy spectra, there is a large number of papers
developing the traditional methods of extracting information from the observations
focusing on the determination of stellar age, metallicity and targeted
abundance ratios \citep[see,  e.g.,][]{JJG:93,Wo:94b,Trager:00,Thomas:05,Gallazzi:05,Graves:09,FLB:13}.
These models range from methods based on the concept of a
simple stellar population (uniquely defined by a stellar initial mass
function and age, along with a fixed chemical composition) to complex
mixtures spanning a range of those parameters. The most detailed
models invoke comparisons of targeted line strengths (such as the Lick
system, and variations thereof) or full spectral fitting
\citep[e.g.][]{SL:05,STECMAP,pPXF:12}. However,
all these methods are hampered by the so-called age-metallicity-dust 
degeneracy whereby changes in age can mimic the effects on the
photometric and spectroscopic information of a change in chemical
composition \citep{Wo:94}. Most notably, this degeneracy is present at
all scales regarding spectral resolution, with similar behaviour in
broadband photometry and spectroscopy \citep{FCS:99}. It is in fact a
major paradox that the amount of information that can be gathered
either from a few broadband colours, or from thousands of fluxes in a
spectrum is rather comparable, reflecting the strong correlation of
spectral features due to the underlying astrophysics.

Alternatively to model comparison methods, other works adopt data-driven 
multivariate techniques aimed at disentangling the information via
principal component analysis \citep{Ronen:99,HCGPCA:06},
factor analysis \citep{Nolan:07}, independent component
analysis \citep{Kaban:05}, Fisher information
matrices \citep{HJL:00}, Bayesian latent variable
analysis \citep{Nolan:06} or the Information
Bottleneck \citep{IB:01,IF:12}. The central tenet of these methods
is that the spectra of galaxies represent superpositions of more
basic units -- eventually individual stellar spectra -- so
that a statistical analysis would reveal the basis spectra
that provide a more fundamental set to decipher the formation and
evolution of galaxies. For instance, it was found that a basis set
comprising just a few 
components could accurately describe sets of thousands of galaxies
\citep{Madgwick:02}, and these components were mostly dependent on age, metallicity
or recent star formation \citep{Madgwick:03,Rogers:07,Wild:14}.
This drastic dimensional reduction can be exploited, for instance, to
speed up the retrieval of star formation histories from spectra
\citep[e.g.,][]{MOPED:03,VESPA:07}. However, while these methods only
depend on the input data and are not affected by potential systematics
of the synthesis models, the results are rather similar in the sense
that only very generic constraints can be produced regarding the star
formation histories. More recently, similar results have been obtained
with deep learning methods based on convolutional neural networks
\citep[see, e.g.,][]{Lovell:19,Portillo:20,CLing:21,Teimoorinia:22}, where the algorithms
are only capable of robustly deriving a few, general properties from
galaxy spectra.

This paper performs a novel analysis based on the interpretation of
galaxy spectra as an information carrier, by redefining a spectrum as
a probability distribution function, with its associated entropy,
following the standard approach of information theory. Entropy encodes
the amount of 'surprise' in the outcome of a given system.  In
spectra, this can be naturally defined when the acquisition of a
spectrum is interpreted as a photon counting experiment. In this case,
the simplest scenarios are represented by a laser beam and a white
light source. In the former, the entropy will be very low, given its
highly monochromatic nature. In contrast, 'white' light will produce
maximum entropy.  Throughout this paper we often link entropy to
``information content'', as defined in 
\citet{Shannon:53}.  This is the most fundamental way of assessing
information, as in, e.g., determining the number of bits needed for a
full representation of the data. Subsequent definitions of information
content focus instead on cross-entropy \citep[see,
e.g.][]{IB:01,MacKay:03} involving an additional set of data or a
classification scheme.

Note that the concept of information as number
of bits needed to encode the spectra is intimately linked to the
extraction of stellar population parameters, although in a non-trivial
way: low information would imply either a highly low-dimensional
representation of the problem, or deep entanglement that produces
strong degeneracies that makes the extraction of the underlying
parameters very challenging. As an example, the spectrum of a
blackbody radiator depends on a single parameter (temperature), thus
allowing for a minimal representation. In stellar populations, one can
envision a large number of fundamental parameters (age and chemical
composition distribution of the stars, initial mass function,
etc.). This paper explores whether the information content somehow
reflects this level of complexity or whether the degeneracies result
in a much simpler scenario regarding spectra as probability
distributions.

Extraction of population parameters from galaxy
spectra depends on diverse practical issues: spectral resolution,
modeling line widths, etc, that vary between data sets and plague
model comparisons. Our proposal to use entropy content is meant to
provide a fundamental procedure to assess the information
content of galaxy spectra and the ultimate limits attainable for
inferring the parametric nature of the stellar populations.
We compare the results from models of population synthesis with actual
galaxies from the Sloan Digital Sky Survey (SDSS). 
We stress that our aim here is not
to suggest new spectral regions or to supersede the traditional
methodology based on model comparisons, but to explore the inverse problem
of extracting stellar population parameters from galaxy spectra from
a standpoint based on information theory. As we will see, the 
analysis illustrates the
inherent entanglement that explains why highly sophisticated
algorithms, such as those based on Deep Learning, can only produce  generic 
descriptions of the stellar content. We emphasize that any
study aimed at the derivation of detailed properties
about the stellar populations from galaxy spectra 
will be subject to this fundamental limit concerning information
content.

The paper has a concise structure, and is organised as follows: the
basic methodology based on entropy is presented in Sec.~2, motivated
by the use of models of stellar population synthesis. The models
produce interesting results regarding the restricted region where
information (negentropy) is substantial, and motivates the selection
of a set of targeted spectral windows. The analysis then turns to the
spectra of galaxies from the Sloan Digital Sky Survey (SDSS) to
confirm whether the suggestions from the models are applicable to real
data, presented in Sec.~3. Finally, our conclusions are given in
Sec.~4 followed by an epilogue that proposes a similar outlook with a
more standard treatment based on covariance.

\begin{figure}
  \centering
  \includegraphics[width=85mm]{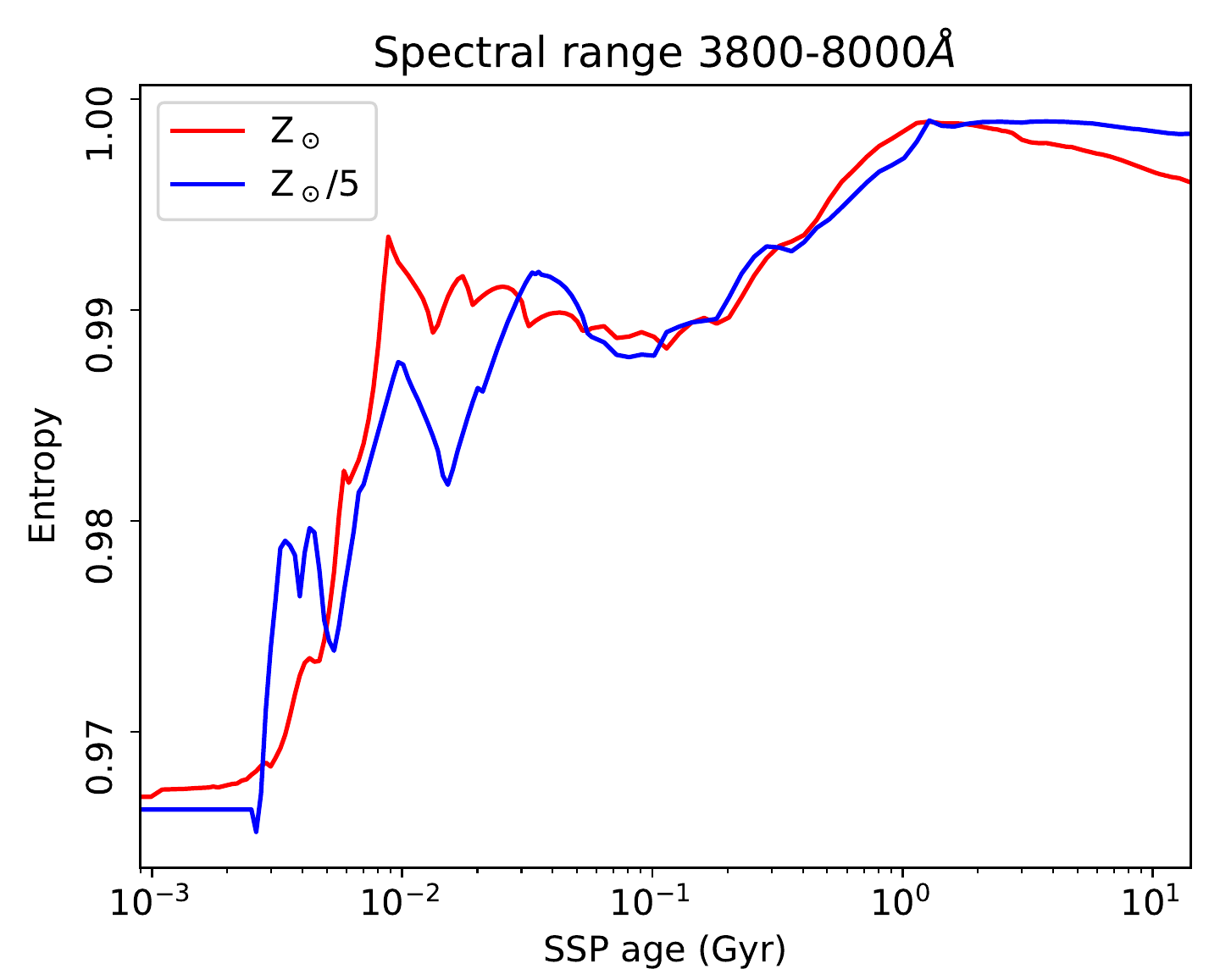}
  \caption{Evolution of entropy with respect to age for a set of simple
    stellar populations (SSPs) from the models of \citet{BC03},
    corresponding to two values of metallicity, as labelled. The entropy,
    defined with the standard definition, eq.~\ref{eq:H1}, 
    is measured within the 3,500--8,000\AA\ spectral range, and is
    defined such that H=1 corresponds to a maximally uninformative case
    (i.e. $p(\lambda)$=constant).}
  \label{fig:Ht}
\end{figure}

\section{The entropy of galaxy spectra}
\label{Sec:Hglx}

The best way to illustrate the connection between spectroscopy and
information theory is to consider
the measurement of a spectrum as a photon
counting process. The specific
flux density can then be interpreted as  a conditional probability distribution. For a
galaxy $g$, the flux at wavelength $\lambda$ is thus:
\begin{equation}
\Phi_g(\lambda) \quad \Longrightarrow \quad p(\lambda | g).
\end{equation}
This formalism was introduced by  \citet{IB:01} to motivate
a classification algorithm based on the so-called ``information bottleneck'',
whereby a sample of galaxy spectra is subject to an agglomerative
binning procedure based on the evolution of entropy, as
spectra are progressively binned into classes. The mutual information
contained in the class representation of the original set serves as a
target to achieve a description of the data with the maximum amount of
information encoded into the smallest number of classes.

In this paper, we do not follow this approach. However,
this interpretation allows us to exploit the concept of
entropy in galaxy spectra as a fundamental way to quantify the way
information is stored across the spectral window. The Shannon
definition of entropy for a probability distribution is \citep{SW:75}:
\begin{equation}
H(\Phi) \equiv -\sum_i p(\lambda_i)\log p(\lambda_i).
\label{eq:H1}
\end{equation}
Hereafter, we denote the entropy with the letter $H$, following
standard notation.  This methodology allows us, for instance, to
quickly identify the language of a given text, by use of the frequency
of the different letters of the alphabet, leading to the definition of
the entropy of a  language \citep{Shannon:51}. In our case, the goal is to
identify the most likely star formation history corresponding to a
given spectrum. In contrast with the traditional methods based on
spectral fitting or targeted line strengths \citep[see,
  e.g.,][]{Walcher:11}, we focus here on the information content as
quantified by the entropy.

\begin{figure}
  \centering
  \includegraphics[width=85mm]{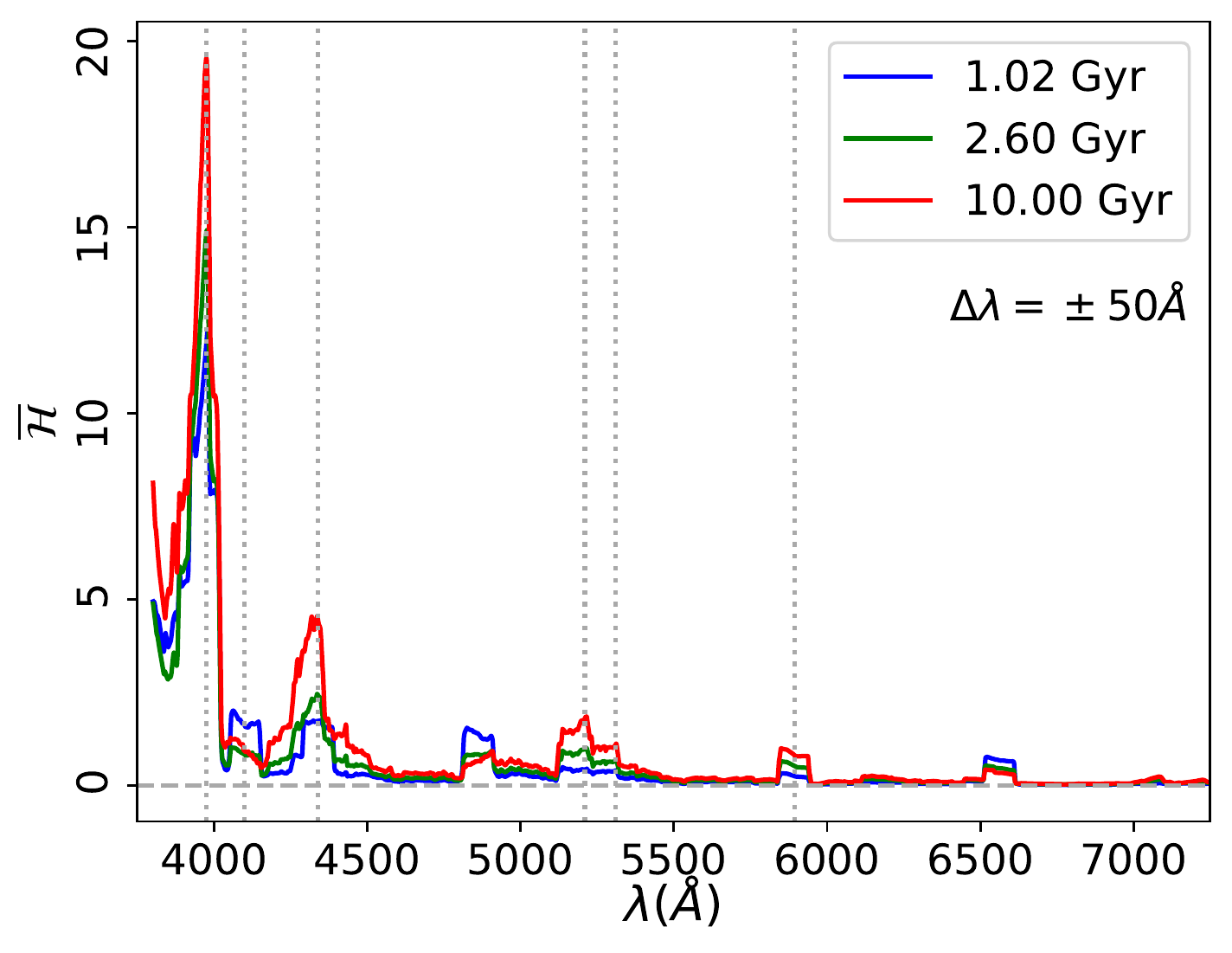}
  \caption{Negentropy spectra ($\overline{\mathcal{H}}(\lambda)$, see eq.~\ref{eq:H2}), 
    corresponding to three SSPs from the
    models of \citet{BC03} at solar metallicity, with age as labelled. 
    They have been computed by measuring the entropy in windows of
    $\delta\lambda$=100\AA. The vertical dotted lines mark the wavelengths chosen for the
    reduced dimensional analysis of SDSS data based on maximum entropy
    variations.}
  \label{fig:Hlam}
\end{figure}

We begin with a simple test determining $H$ in a set
of well-known and tested population synthesis models. These models incorporate our knowledge
of the formation and evolution of stars, as well as the radiative properties
of stellar atmospheres that ultimately give rise to spectra
\citep{Conroy:13}. These models
are typically produced for a reduced set of variables, mainly stellar initial
mass function, age, 
total chemical composition (metallicity), and possibly a number of non-solar
abundance ratios, most notably [Mg/Fe]. At present, state of the art
models rely on a reduced set of libraries of stellar spectra
\citep[see, e.g.,][]{STELIB,Coelho:05,MILES}, so that -- from
the pure point of view of
information theory -- any synthetic model, no matter how complex, is
always defined by a
linear combination of a relatively reduced set of stellar spectra, of order
$10^3$. This is the reason why information-based models can easily
discriminate between synthetic and ``real'' spectra, as shown
in \citet{IB:01} where the information content from observed 2dFGRS
spectra was found to be much higher than the one derived from
theoretical galaxy formation models that incorporated population
synthesis\footnote{Also noting that noise -- inhererent to any
observational data -- will affect the information content.} (see figure~3 in that paper).

\begin{figure}
  \centering
  \includegraphics[width=85mm]{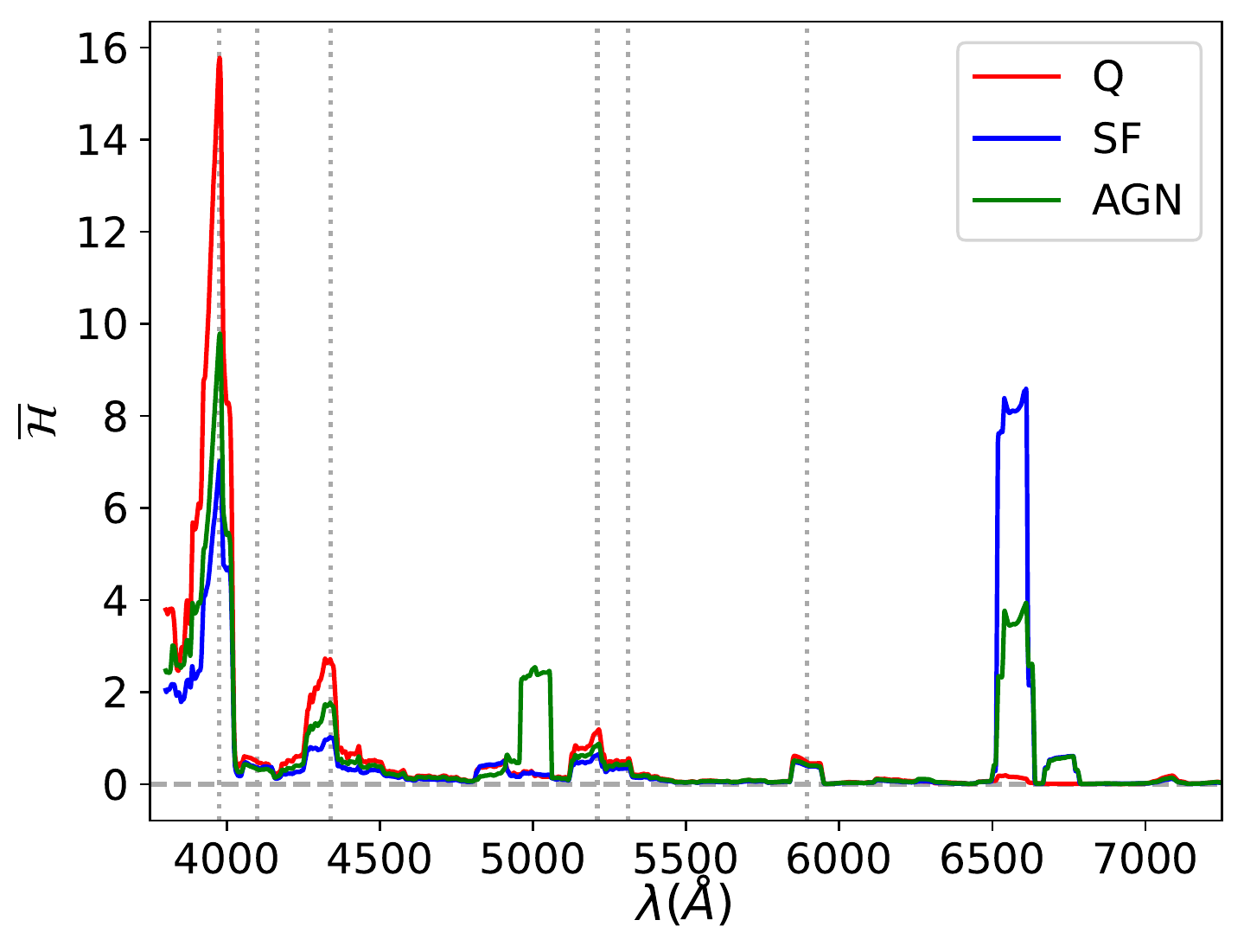}
  \caption{Negentropy spectra of SDSS galaxies, equivalent to the
    one shown for synthetic spectra in Fig.~\ref{fig:Hlam}. Here we plot 
    separately quiescent (Q, red), star-forming (SF, blue) and AGN
    (green) galaxies based on their nebular emission properties,
    following the standard BPT classification (see text for
    details). The vertical dotted lines mark the wavelengths chosen for
    the reduced dimensional analysis. Note the substantial contribution
    to the entropy from emission lines in SF and AGN spectra. These lines
    correspond to the diffuse gas component, not the stellar populations,
    and thus are omitted in our analysis.}
  \label{fig:HlamSDSS}
\end{figure}

Fig.~\ref{fig:Ht} shows the entropy defined within a relatively wide
spectral window, between 3,500 and 8,000\AA, normalized such that a
totally uninformative scenario -- i.e. a constant $p(\lambda)$ in this
interval -- corresponds to $H=1$, and the extreme case of a perfectly
monochromatic signal, i.e. $p(\lambda)=\delta(\lambda-\lambda_0)$ has
$H=0$. The simple stellar population models (SSPs) of \citet{BC03} have
been used here to explore the variation with respect to stellar age
(horizontal axis), with two choices of metallicity, as labelled.
We adopt the \citet{Chabrier:03} stellar initial mass function in this
analysis, but the differences are minimal for other reasonable
choices. 
From an information theory point of view, the first salient feature
of galaxy spectra is the closeness to a fully uninformative scenario: the
deviation from the maximum entropy case is less than one part in $\sim$30.
However, this is caused by the large spectral interval chosen and by the fact
that, aside from absorption features (that usually amount to less than a few Angstrom
in equivalent width), the continuum is a fairly smooth function of
wavelength, with
a relatively shallow gradient. Moreover, the trend found with respect to age
stems from this: the younger populations are dominated by hotter stars with
steeper, and blue, continua, whereas older populations have a much
shallower wavelength dependence, except for the prominent 4,000\AA\ break.

In order to be able to discriminate better among the models, we
measure the entropy within a narrower spectral window. This is the
equivalent of defining individual probability distributions within
prescribed wavelength intervals: $p(\lambda | [\lambda_1,\lambda_2])$,
where only photons with wavelength between $\lambda_1$ and $\lambda_2$
are considered in the definition of the probability distribution
function. We can also do this experiment adopting a running interval,
of width $\Delta\lambda$, as we traverse the optical window, defining
an ``entropy spectrum'', $H_{\Delta\lambda}(\lambda)$.  The choice of
$\Delta\lambda$ is important: too wide and we wash out all
information, too narrow and the result becomes prohibitively dependent
on the signal to noise ratio or velocity dispersion. At the resolution
of SDSS galaxy spectra, around R$\sim$2,000, \citet{Rogers:10} found
that $\Delta\lambda\sim$50-100\AA\ provides an optimal window, a
result that is equally optimized for stellar spectra in SDSS
\citep{Hawkins:14}. Fig.~\ref{fig:Hlam} shows the ``entropy spectrum''
of a few synthetic populations from the models of \citet{BC03}, using
a running interval of width $\Delta\lambda$=100\AA. As comparison, we
show the corresponding entropy spectrum for a sample of real galaxy
spectra from SDSS in Fig.~\ref{fig:HlamSDSS} -- showing the median of
subsamples classified according to their nebular emission properties
(Q: quiescent, SF: star-forming, AGN: active nucleus).  We will
explore this sample in more detail in \S\ref{Sec:SDSS}.  Hereafter, we
quote estimates of entropy with the following definition:
\begin{equation}
\overline{\cal H}(\lambda)\equiv [1- H_{\Delta\lambda=100A}(\lambda)]\times 1,000,
\label{eq:H2}
\end{equation}
which can be interpreted as negentropy, or information content.  The
maximally uninformative case  corresponds to
$\overline{\cal H}$=0. Note that in Fig.~\ref{fig:Ht} the entropy variations
are, at most of a few per 100.
The number 1,000 in eq.~\ref{eq:H2} is just a scaling factor so that
the values of negentropy are given in numbers of order $\sim$10.

Fig.~\ref{fig:Hlam} provides insight into the information encoded in galaxy
spectra. At redder wavelengths, the spectrum becomes less informative,
as it is dominated by a continuum with relatively weak absorption
features, except for some prominent regions, as labelled.  The
strongest contribution comes from the region around 4,000\AA, where
the pile up of spectral lines produces a prominent break, with the
entropy reaching its lowest value. While minute, targeted line
strengths can be exploited to constrain stellar population
properties \citep[see, e.g.,][]{FLB:13}, other methods based on full
spectral fitting, or machine learning algorithms -- where the full
spectra are directly input into the algorithm -- will be subject to
this problem.

Note how this blind approach to the information content of spectra produces the
standard features targeted for the analysis of stellar populations: the Balmer
regions at 4,100\AA\ (H$\delta$); 4,340\AA\ (H$\gamma$); 4,862\AA\ (H$\beta$)
and 6,563\AA\ (H$\alpha$) are more prominent in younger populations around 1\,Gyr,
a well-known result caused by the dominance of A-type stars to the net luminosity
budget in populations with that age. i.e. when that stellar type becomes the
main sequence turn-off.
Older populations produce conspicuous features around 4,300\AA\ (G band)
and 5,100-5,400\AA, where a number of Mg and Fe absorption lines are very prominent.
The optical Na absorption at 5,890\AA\ is also present in the entropy spectra.
These  features have been intensively explored in galaxy spectra
for decades, so their presence is not surprising. However, our analysis 
confirms that these regions are the ones that carry most of the
information -- as in negentropy -- from a pure information theory
viewpoint. This result is especially relevant to Deep Learning
algorithms, where the complex ``machines'' critically depend on the
information content of the spectra to classify or learn about the 
underlying components.  In essence, machine learning methods rely on a
quantitative assessment, based on some figure of merit, whose
maximisation leads either to a decision (in a classification algorithm),  
or to a set of best-fit parameters (in a regression). 
The success of any ML method -- either supervised or unsupervised -- will
ultimately depend on how this figure of merit can discriminate among
the basis of sources. These sources correspond here to the generic base set of simple
stellar populations that describe any star formation history.  Blind
source separation methods also operate
on the same basis \citep[see, e.g.,][]{ICA}, namely that the original
sources that produce the final data carry enough information to be
told apart from the observed mixtures.  Whilst we do not
optimally fit specific line strengths given a set of models, the
analysis of entropy marks an intrinsic limit regarding the ability of
any such method to unambiguously extract the details of the underlying
stellar populations from spectroscopy.

\begin{figure}
  \centering
  \includegraphics[width=90mm]{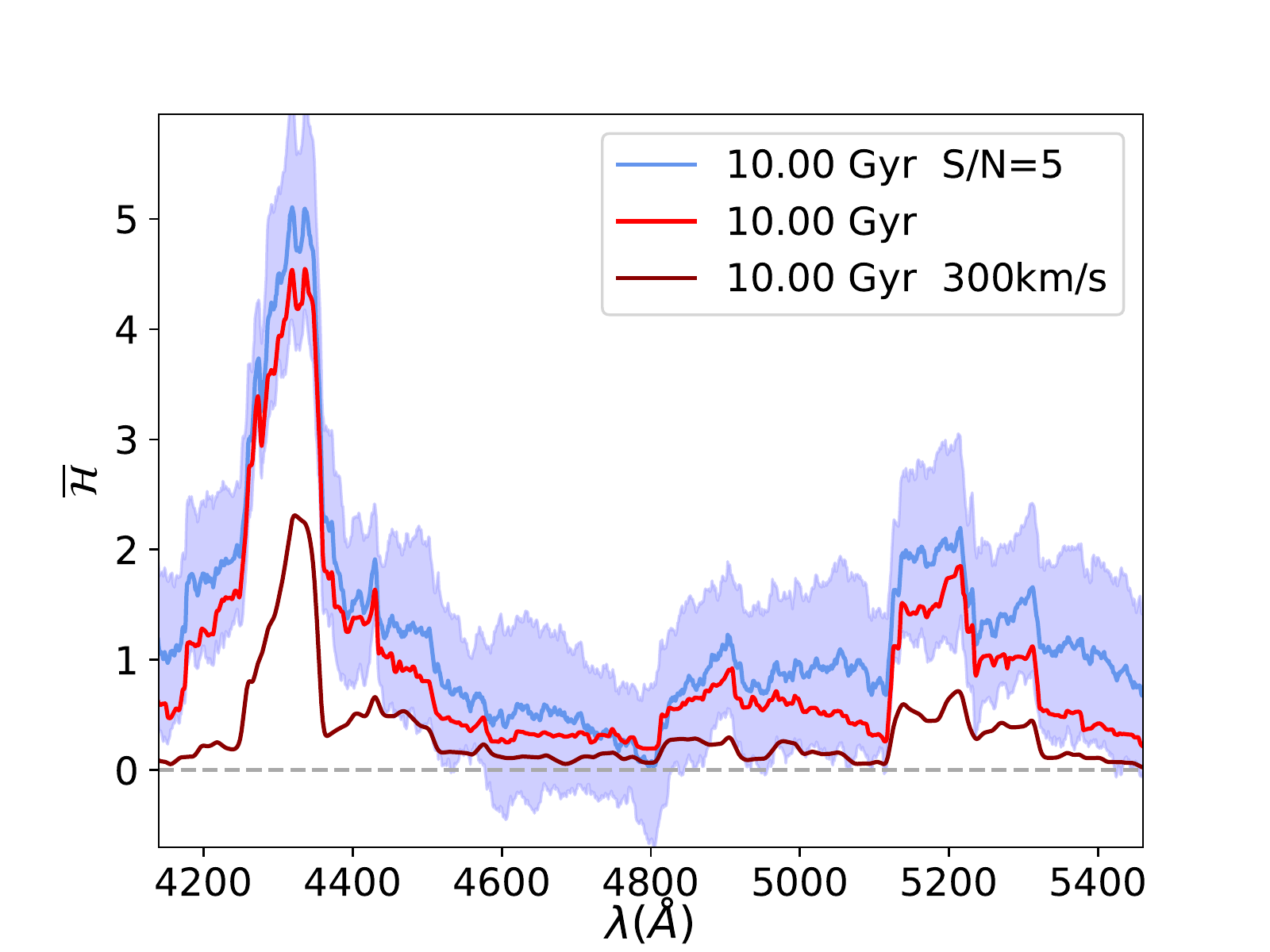}
  \caption{Variation of the negentropy spectrum with respect
    to velocity dispersion and noise. A test case
    is shown for a 10\,Gyr SSP at solar metallicity,
    within a relatively narrow spectral window for ease
    of visualization. 
    The red line represents the noiseless reference
    at ``zero velocity dispersion''.}
  \label{fig:Hvd}
\end{figure}

This paper is focused on how information is stored, and can be
retrieved, from galaxy spectra in the most fundamental way based on
entropy.  The results from the models justify a simplification of the
methodology by selecting a small set of targeted regions that correspond to the
highest values of negentropy. Please note we avoid on purpose H$\beta$
and H$\alpha$, as they are strongly affected by emission from the gas
component, and therefore bias any analysis focused on stellar
populations. Traditional studies typically remove these lines from the
analysis or perform detailed subtraction of the emission from the gas,
but this introduces a substantial dependence on the models, and leads
to systematics. Moreover, the emission lines trace ``instantaneous'' properties of the galaxy,
namely the ionisation of the diffuse gas by young stars, AGN activity or shocks. The analysis
of populations focuses, rather, on the so-called ``fossil record'', i.e. the star formation
activity throughout the whole formation history of the galaxy.
Entropy in each of these six intervals is
measured within the $\Delta\lambda$=100\AA\ window.
The vertical dotted lines mark those regions -- adopted below for the
analysis of real galaxy spectra from SDSS (see Sec.~\ref{Sec:SDSS}).
While Fig.~\ref{fig:Hlam} is shown for a few choices of stellar age and solar metallicity,
the models feature similar behaviour at different ages and chemical composition.
Therefore we propose a simplified description of each
galaxy spectrum with only six ``coordinates'', that represent the regions
that carry the maximum amount of variation in their entropy.

It is worth noting that the interpretation of a galaxy spectrum as a probability
distribution of the energy of the incoming photons is affected by two observational
pitfalls: the signal-to-noise ratio and the effective spectral resolution. The former
depends on the efficiency of the observing apparatus and the exposure time, whereas the
latter depends both on the spectrograph as well as the galaxy under scrutinity, as the
distribution of stellar orbits impose an effective ``kinematic kernel'' that broadens
-- via Doppler shifts -- all
spectral features. Fig.~\ref{fig:Hvd} illustrates the effect on the entropy, with
the noiseless case at the fiducial resolution of the stellar models shown as a
red line. A reasonable S/N still produces manageable results, with the shaded region
spanning the range of entropy produced by a bootstrap that adds Gaussian noise to
the model spectra, corresponding to S/N=5 per \AA. A lower spectral resolution -- here shown
by producing the equivalent observation of a (massive) galaxy with
velocity dispersion of 300\,km/s (dark red) degrades the entropy quite
dramatically. As expected, lower resolution (or higher velocity
dispersion) produces smoother spectra, with less defined absorption
features, approaching a homogeneous probability distribution, and
therefore tending towards maximum entropy, i.e.  minimum
information. For those reasons, the analysis presented in the next
section -- comprising actual galaxy spectra -- will constrain the data
to high S/N and relatively low velocity dispersion.

\begin{figure}
  \centering
  \includegraphics[width=85mm]{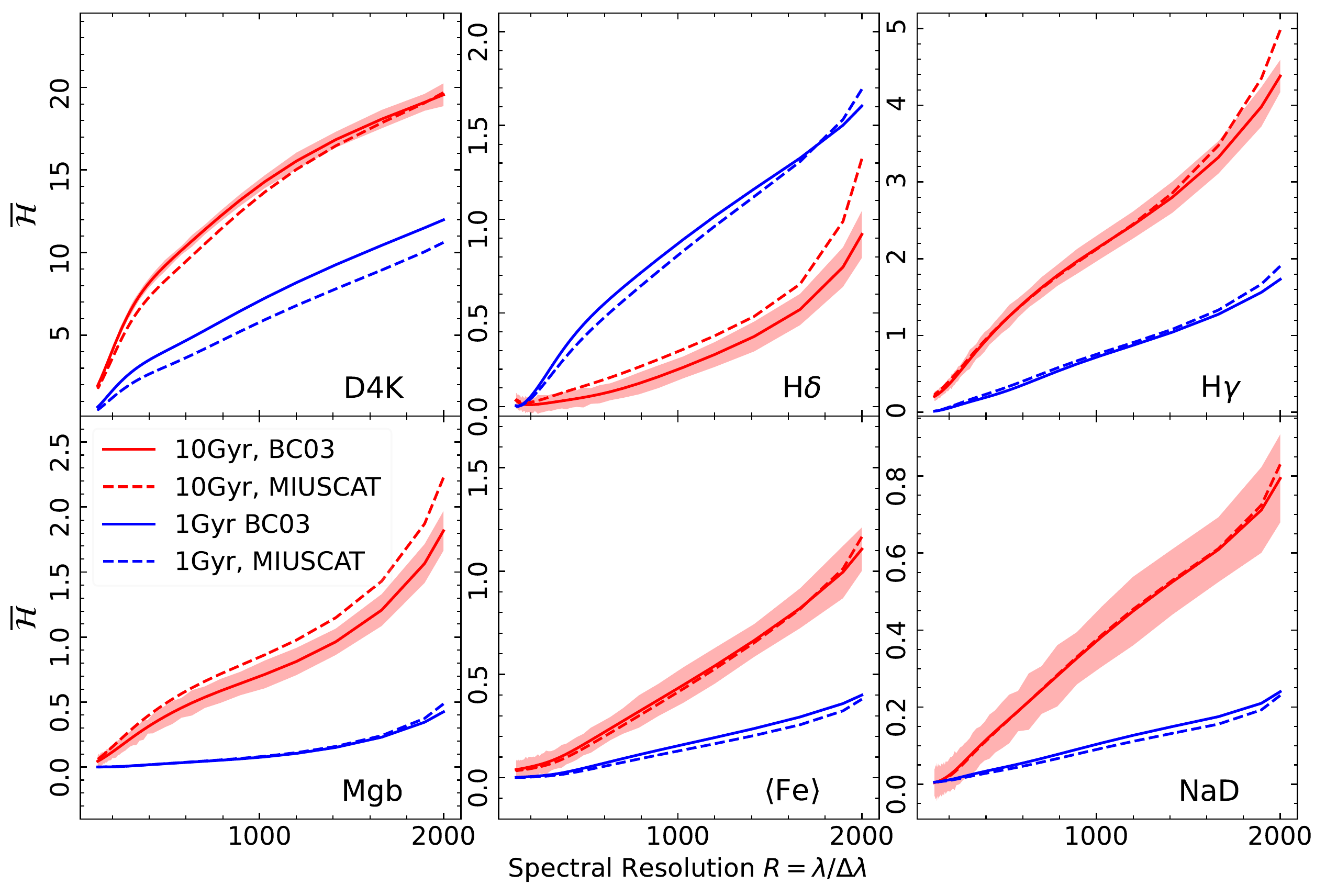}
  \caption{Variation of the information content of galaxy spectra with
    respect to the effective spectral resolution (horizontal axes),
    shown in the six spectral windows targeted in this paper. The blue
    (red) lines represent a young (old) population at solar metallicity
    and solar abundance ratios from the \citep[BC03,][solid]{BC03} and
    \citep[MIUSCAT,][dashed]{MIUSCAT} population synthesis models,
    as labelled. The red
    shaded region shows, for reference, the expected 1\,$\sigma$
    uncertainty at a S/N of 20\,\AA$^{-1}$.}
  \label{fig:HRes}
\end{figure}

Fig.~\ref{fig:HRes} shows how entropy changes with respect to spectral
resolution within the six features defined here. The results are shown
for two  population synthesis models at two different stellar ages
(both with solar chemical abundance), as labelled. The \citet{BC03}
models are shown as solid lines, and the MIUSCAT models \citep{MIUSCAT}
are plotted with dashed lines. We emphasize that these models are
independent, and even adopt different stellar libraries.
Note the trends are equivalent
in both sets of models, although there are some systematic variations 
caused by the different choice of model prescriptions and stellar libraries. 
For reference, we include an estimate of the 1\,$\sigma$ uncertainties if the
spectra have a S/N of 20\,\AA$^{-1}$, as a shaded region for the 10\,Gyr BC03 model.
Noting that R$\sim$2,000 is an optimal resolution for
galaxy spectra given the typical velocity dispersion of the stellar
component, we find that the information content -- defined as the
entropy -- gets halved at R$\sim$500-1000 (varying with the spectral
index). 
The left part of the horizontal axis corresponds to the
resolution expected in slitless grism spectroscopy or medium band
photometry, where the discriminatory power of spectral indices is
reduced -- in this case, the continuum is used instead, to
constrain the population parameters \citep[see, e.g.][]{PEARS:09,SHARDS,LDG:15}.

\section{The entropy content of SDSS galaxy spectra}
\label{Sec:SDSS}

The trends presented in the previous section have been obtained with synthetic models of
stellar populations. In this section, we turn the entropy analysis to 
a set of high quality galaxy spectra from the Sloan Digital Sky Survey
\citep[SDSS,][]{York:00}.  We use the original, legacy, survey that consists of
single fibre spectroscopy at resolution ${\cal R}\sim$2,000. From the
original dataset  (we retrieve the data from their Data Release 16,
\citealt{DR16}), we select a sample of 76,569 spectra with a high S/N
($\gtrsim$10\,\AA$^{-1}$, measured in the $r$ band),
constrained in velocity dispersion between
$\sigma$=100 and 150\,km/s. It is a well-known fact that the stellar
velocity dispersion -- a tracer of the gravitational potential --
correlates strongly with population properties \citep[e.g.,][]{Bernardi:03,SAMIGrad}.
By choosing a
relatively reduced range in $\sigma$, we aim at simplifying the
working sample. Our goal in this paper is to explore how well entropy
can disentangle subtle differences in stellar populations, so a
focused data set is preferred.
The median redshift is z=0.079 with a 95\% interval
$\Delta$z=[0.054,0.099]. The spectra are corrected for foreground dust
contamination using the dust extinction law of \citet{Fitz:99},
are brought to the rest-frame, and are resampled
to a common wavelength grid with 1\AA\ spacing.
In each spectrum, bad pixels flagged as problematic by the SDSS team
-- that represent a very small fraction in these high S/N data -- 
were replaced by their best model fits, also provided within the data
structure of the SDSS spectra. 

Fig.~\ref{fig:HlamSDSS} shows the negentropy spectrum of SDSS data,
i.e. the measurement of $\overline{\mathcal{H}}(\lambda)$
(eq.~\ref{eq:H2}) with a running $\Delta\lambda$=100\AA\ interval, as
in Fig.~\ref{fig:Hlam}, which was presented for a few simple stellar
populations from the synthetic models of \citet{BC03}.  The spectra
are median-stacked into three classes, defined according to nebular
emission as quiescent (Q, red), star-forming (SF, blue) and AGN
(green). These three classes represent a fundamental transition during
the evolution of galaxies, from an initial star-forming stage (in the
Blue Cloud) to passive evolution in a quiescent phase (in the Red
Sequence). AGN activity appears to dominate the transitioning stage
(mostly populating the intermediate region, termed the Green
Valley). These regions stem from the inherent bimodality of galaxy
properties \citep{Strateva:01, Salim:14}.  Therefore, by splitting the
sample according to this classification scheme, we can assess how such
a transition affects the entropy of galaxy spectra.
The classification follows the standard BPT diagram \citep{BPT}, and is
taken from the official galspecExtra SDSS catalogue \citep{MPA_JHU},
where we include an additional constraint on the equivalent width of
H$\alpha$ emission to select quiescent galaxies \citep[e.g.][]{CF:11}
-- as the bpt=$-$1 flag only refers to galaxies whose spectra cannot
be mapped on the standard BPT diagram.  The real spectra show a
similar behaviour to the models, except at the location of prominent
emission lines such as H$\beta$, [O{\sc III}], H$\alpha$+[N{\sc II}],
and [S{\sc II}].

\begin{figure}
  \centering
  \includegraphics[width=80mm]{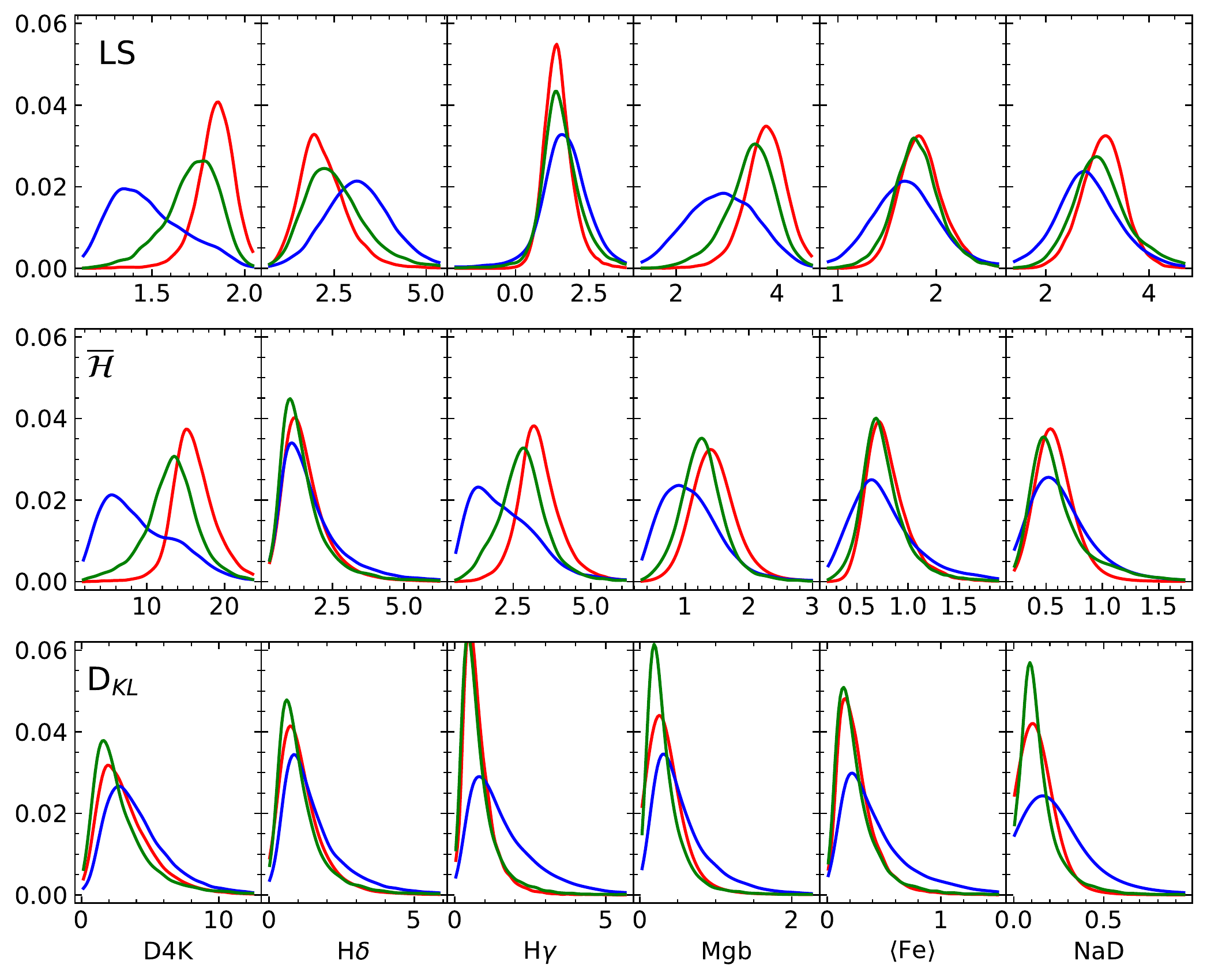}
  \caption{Distribution of line strength equivalent widths (top), information
    content based on negentropy ($\overline{\cal H}$, middle)
    and Kullback-Leibler divergence (D$_{\rm KL}$, bottom) for 
    a set of galaxy spectra from SDSS. The sample is subdivided into
    star-forming (blue), quiescent (red) and AGN (green) galaxies
    based on the nebular emission properties from the standard BPT
    classification scheme (see text for details). The histograms
    can be interpreted as probability distributions, and are plotted
    with a Gaussian kernel density estimator. 
  }
  \label{fig:SDSSHist}
\end{figure}

We focus on the most relevant features of the negentropy spectra, as
found in the models, namely 4000\AA\ break strength (termed $D4K$),
two Balmer absorption lines (H$\gamma$, H$\delta$) and the strongly
metallicity dependent indices Mgb, $\langle$Fe$\rangle$ and NaD
\citep[see, e.g.][]{WO:97,Trager:98,Balogh:99}. 
We will compare the results from those six spectral windows in three
complementary ways: 1) line strengths, measuring the equivalent
width in those windows, following the methodology of \citealt{Rogers:10});
2) negentropy,
as defined above; 3) relative entropy, also termed Kullback-Leibler divergence,
D$_{\rm  KL}$, \citealt{Dkl} (bottom).  The colour coding separates
subsamples based on nebular emission (Q: red, SF: blue, AGN: green).
The  definition of relative entropy rests on the idea of the information content of
a probability distribution, $p(\lambda)$, with respect to another one
that acts as baseline, $q(\lambda)$. In this case, entropy refers to
the amount of information relative to the baseline:
\begin{equation}
D_{\rm KL} \equiv -\sum_i p(\lambda_i)\log \left[p(\lambda_i)/q(\lambda_i)\right].
\label{eq:Dkl}
\end{equation}
Note that when $q(\lambda)$ is constant (i.e. a totally uninformative
baseline), we recover Shannon entropy. Here we use as reference
the median of all spectra, taking the galaxies from the whole
sample, regardless of nebular emission (i.e. including Q, SF and AGN), 
to define the median $q(\lambda)$. Therefore, relative entropy in this paper
is a measure of departure from the overall behaviour of the general sample.
Fig.~\ref{fig:SDSSHist} shows the distribution of the six spectral
windows in the three different representations of the spectral information:
Line strengths (LS, top);
negentropy ($\overline{\cal H}$, middle) and relative entropy
(D$_{\rm KL}$, bottom). 

The most significant difference between SF, AGN and Q subsamples in
Fig.~\ref{fig:SDSSHist} is obtained for D4K, and the SF subset is the
one with the widest range of values in all cases. Comparing the
standard line strengths with negentropy, we find some analogous
results (D4K, Mgb, $\langle$Fe$\rangle$), whereas Balmer absorption
behaves differently: while the line strength analysis finds a
larger variation regarding nebular activity in H$\delta$, 
negentropy discriminates better with H$\gamma$. In contrast, relative
entropy does not produce such clear segregation, with only a more
extended tail in the distribution of SF galaxies.
The figure shows that there is no evident trend when looking at any of the
six spectral windows. While the separation of spectra based on
nebular emission is a fundamental discriminator of activity, the
distributions do not reveal substantial differences, neither in LS
nor in the two definitions of entropy. Therefore, in order to explore in more detail the
potential drivers of variance/information, we resort  to further
reduce the dimensionality of parameter space via Principal Component Analysis (PCA).
PCA provides a simple way to visualize the variance of the data by diagonalizing
the corresponding covariance (represented here by a $6\times 6$ matrix).
The eigenvectors, or principal components, are ranked in decreasing order of their
associated variance, and the original data are projected onto these eigenvectors
to produce an alternative representation that is ranked as a function of
variance.

\begin{table}
\caption{Normalized PCA eigenvalues.\label{tab:scree}}
\begin{center}
\begin{tabular}{cccc}
\hline
Rank & \multicolumn{3}{c}{Eigenvalue ratio}\\
 & Line Strength & Entropy & D$_{\rm KL}$\\
\hline
1 & 0.50022 & 0.87499 & 0.75479\\
2 & 0.25449 & 0.06983 & 0.09225\\
3 & 0.11622 & 0.02057 & 0.05972\\
4 & 0.09691 & 0.01687 & 0.04688\\
5 & 0.02616 & 0.01478 & 0.03986\\
6 & 0.00600 & 0.00296 & 0.00650\\
\hline
\end{tabular}
\end{center}
\end{table}

We apply PCA independently to the data from the three methods
described above, finding the eigenvalues shown in
Table~\ref{tab:scree}. They are quoted as a ratio of total variance,
and show that the standard line strength analysis produces the most
mixed distribution, needing up to four components to account for 90\%
of total variance, whereas the first principal component for standard
entropy already carries $\sim$87\% of total variance, with the second
one accounting for around 7\% and the subsequent components featuring
a lower share. Relative entropy also shows a high weight of the first
principal component, accounting for $\sim$75\% of total variance.

\begin{figure}
  \centering
  \includegraphics[width=80mm]{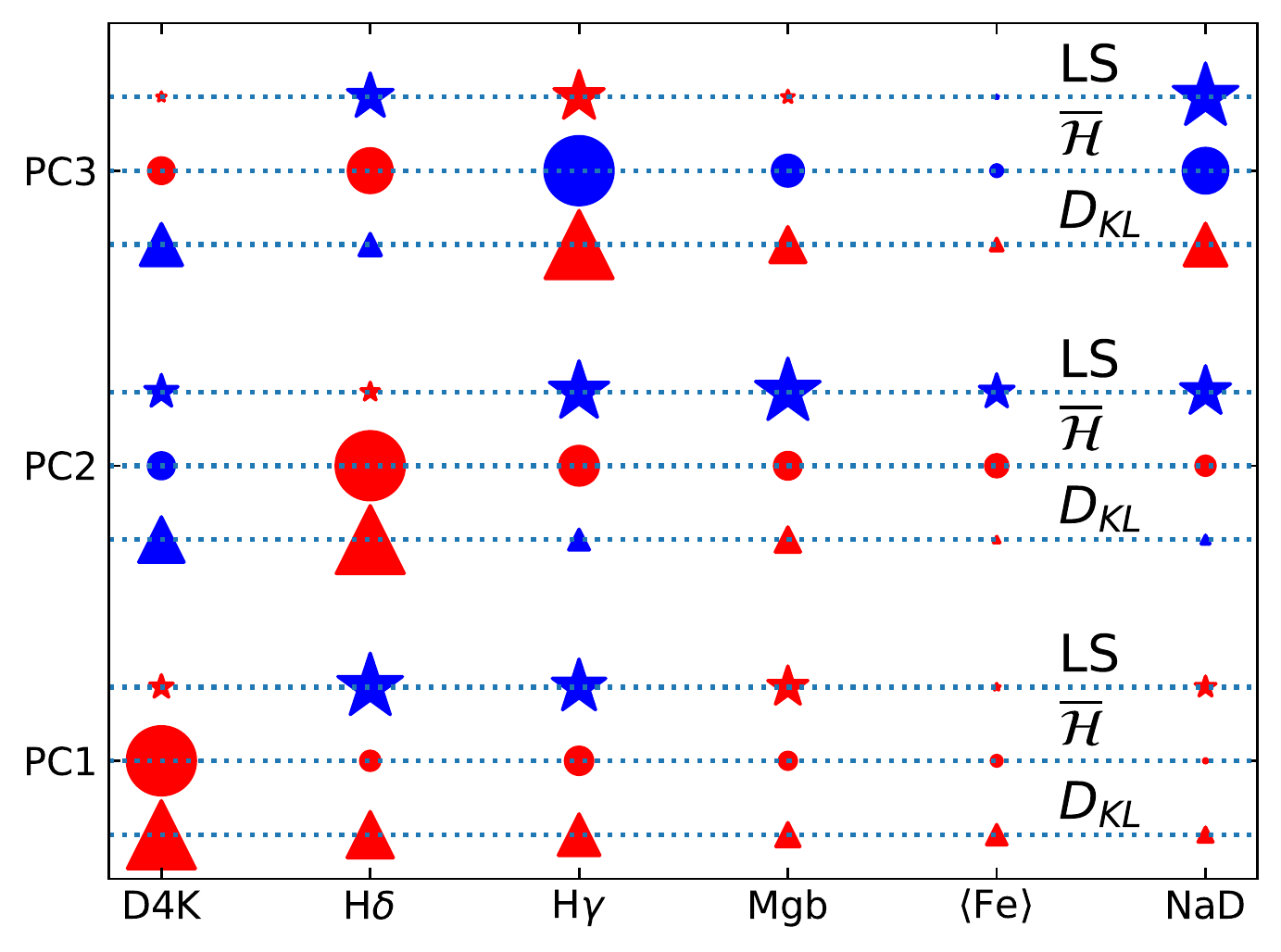}
  \caption{Weights of the first three principal components
    (PC1, PC2, PC3) using as input either line strengths (stars, LS), 
    negentropy (circles, $\overline{\mathcal{H}}$), 
    or relative entropy (triangles, D$_{\rm KL}$), shown for the six
    features adopted to describe the spectra (horizontal axis). The
    symbol size corresponds to the value of the weight, and the
    colour gives the sign as positive (red), or negative (blue).
    Note negentropy gives the simplest interpretation of the variations
    in terms of 4000\AA\ break strength (PC1) and Balmer absorption
    H$\delta$ (PC2).}
  \label{fig:PCA_weight}
\end{figure}

Fig.~\ref{fig:PCA_weight} is a graphic representation of the weights
of the first three principal components, where symbol size denotes the
weight of a given spectral feature, as labelled in the horizontal
axis, and the colour means positive (red) or negative (blue) sign for
each weight.  For reference, these weights are quantified in
Table~\ref{tab:weights}.  The principal components derived from the
standard line strength data (LS) correspond to mixtures of Balmer
absorption (mostly in PC1) and metallicity dependent strengths
(PC2). In contrast, the components corresponding to Shannon
(neg)entropy ($\overline{\mathcal{H}}$) produce a simpler scheme,
where PC1 is mostly represented by D4K and PC2 is dominated by the
information in the Balmer absorption lines (H$\delta$ and
H$\gamma$). The third component -- whose associated variance is only
2\% -- depends mostly on H$\gamma$, with residual dependence on the
other features, most notably H$\delta$ and NaD.  The relative entropy
representation (D$_{\rm KL}$) produces a mixed set of weights, with
PC1 mostly having a strong dependence on D4K and Balmer absorption
(both H$\delta$ and H$\gamma$ being represented). The second component
is dominated by H$\delta$, with a substantial dependence on D4K
(although with negative sign, in contrast to PC1). The third component
mostly relates to H$\gamma$ with some dependence on D4K and
NaD. Surprisingly, the metallicity dependent features (Mgb and
$\langle$Fe$\rangle$) play a minor role in these components, which may
be caused by the strong correlation among spectral indices (see
Epilogue below).  Overall, it is worth emphasizing that the bulk of
the entropy variation found in both representations refer to the
4000\AA\ break strength and H$\delta$.  This type of scheme has been
extensively used in standard analyses of stellar populations in
galaxies \citep[e.g.,][]{Kauff:03} as the 4000\AA\ break indicator is
overall sensitive to the average age and metallicity, whereas
H$\delta$, or H$\gamma$, trace recent episodes of star formation
\citep[see, e.g.][]{Kauff:14, Wu:18}. In this paper, we
retrieve this result purely from the information point of view,
without any reference to the modelling of stellar populations. In
Appendix~\ref{App:LS} we show the actual distribution of line strengths
for the six targeted features, confirming the importance of D4K and
Balmer absorption, with a strong relation to nebular activity, which
acts as a rough proxy of evolution -- i.e. from star-forming to AGN to
quiescence.

\begin{table*}
\caption{First three PCs (normalized)\label{tab:weights}}
\begin{center}
\begin{tabular}{l@{\quad\vline\quad}ccc@{\quad\vline\quad}ccc@{\quad\vline\quad}ccc}
\hline
 & \multicolumn{3}{c@{\vline\quad}}{LS} & \multicolumn{3}{c@{\vline\quad}}{$\overline{\mathcal{H}}$}& \multicolumn{3}{c}{D$_{\rm KL}$}\\
\hline
Region & PC1 & PC2 & PC3 & PC1 & PC2 & PC3 & PC1 & PC2 & PC3\\
\hline
D4K &  0.10581 & $-$0.17878 &  0.01760 & 0.98189 & $-$0.13520 &  0.11992 &  0.84949 & $-$0.40348 & $-$0.32569\\
H$\delta$ & $-$0.77859 &  0.05707 & $-$0.39372 & 0.07952 &  0.92508 &  0.34829 &  0.39536 &  0.90203 & $-$0.09239\\
H$\gamma$ & $-$0.53260 & $-$0.56543 &  0.47513 & 0.15791 &  0.30010 & $-$0.83879 &  0.31989 & $-$0.08616 &  0.84611\\
Mgb &  0.30326 & $-$0.66813 &  0.03245 & 0.06321 &  0.14242 & $-$0.17315 &  0.10898 &  0.12565 &  0.23871\\
$\langle$Fe$\rangle$ &  0.00990 & $-$0.19417 & $-$0.00298 & 0.02505 &  0.09990 & $-$0.02560 &  0.07778 &  0.00882 &  0.02985\\
NaD &  0.08290 & $-$0.40121 & $-$0.78604 & 0.00390 &  0.07500 & $-$0.36071 &  0.04251 & $-$0.01625 &  0.33409\\
\hline
\end{tabular}
\end{center}
\end{table*}

\begin{figure}
  \centering
  \includegraphics[width=80mm]{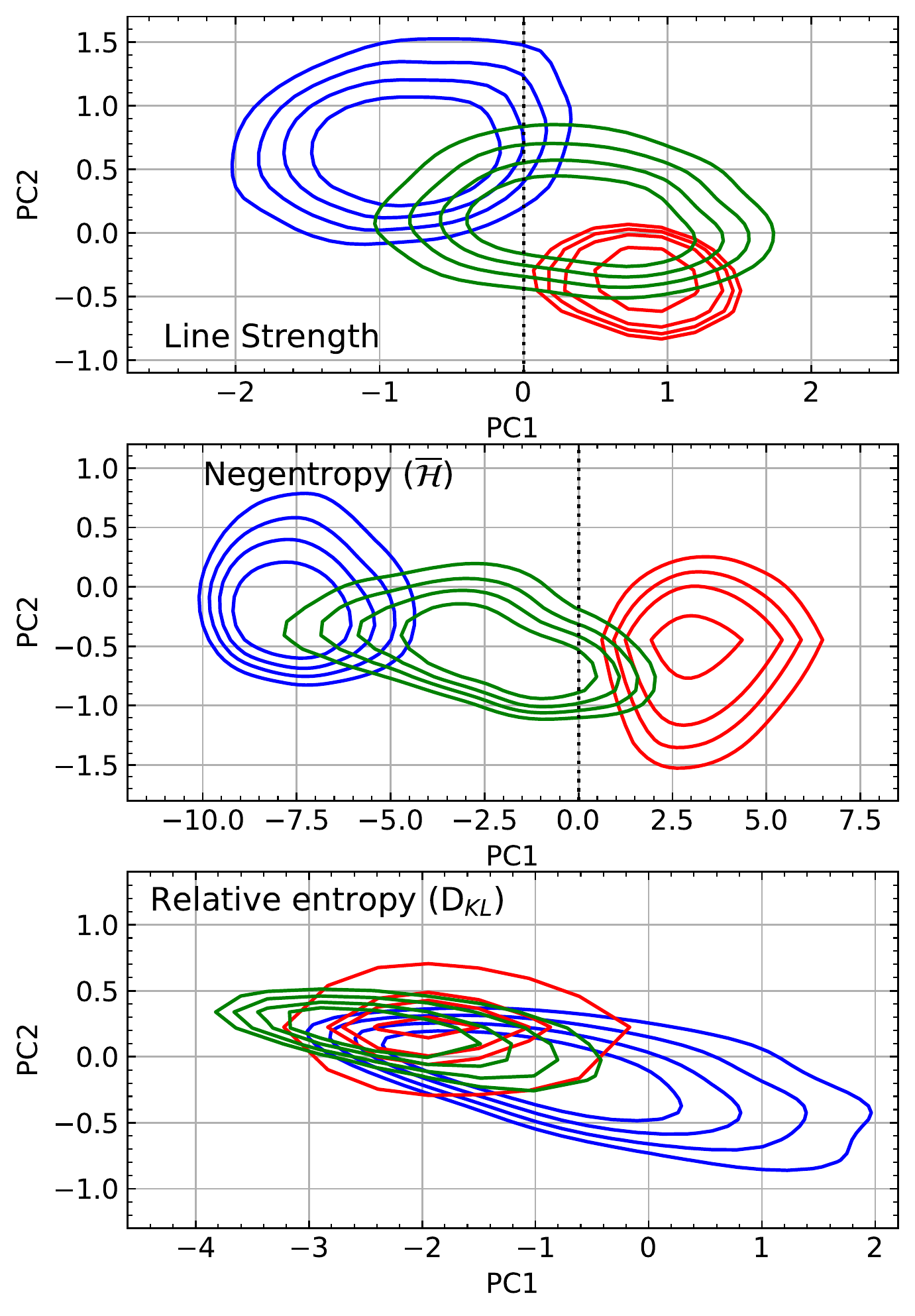}
  \caption{The sample of SDSS spectra is projected
    on to the first two principal components, with
    the distributions shown as contours (corresponding
    to their number density, from 50\% to 80\% of
    each sample, in steps of 10\%. The sample is
    shown separately for star-forming (blue),
    quiescent (red) and AGN (green) spectra.
    The projection for line strength, negentropy
    and relative entropy, is shown in the top, middle, 
    and bottom panels, respectively. Negentropy produces
    the largest discrimination among these with respect
    to a single component (PC1).}
  \label{fig:PCA_conts}
\end{figure}

\begin{figure}
  \centering
  \includegraphics[width=85mm]{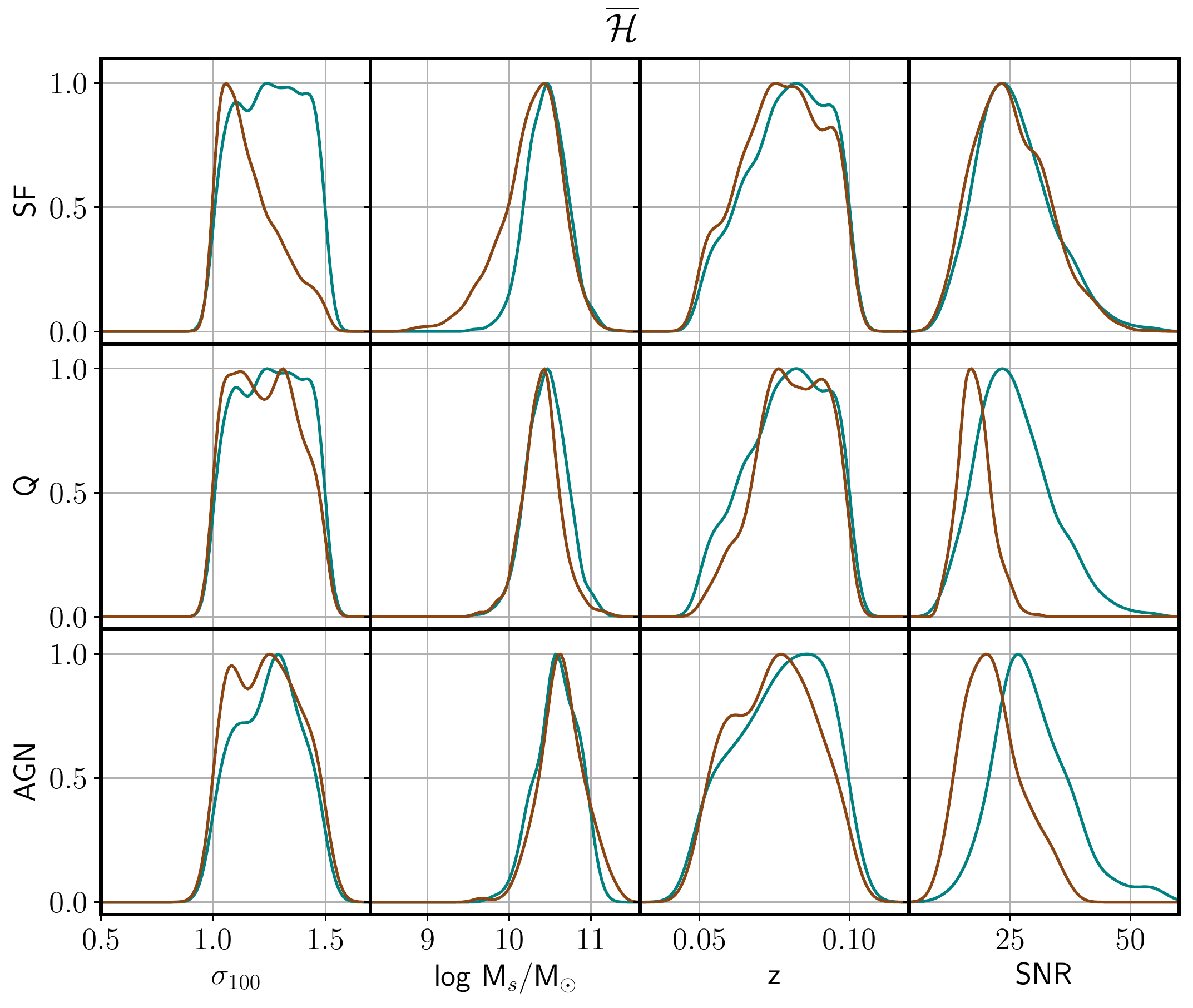}
  \includegraphics[width=85mm]{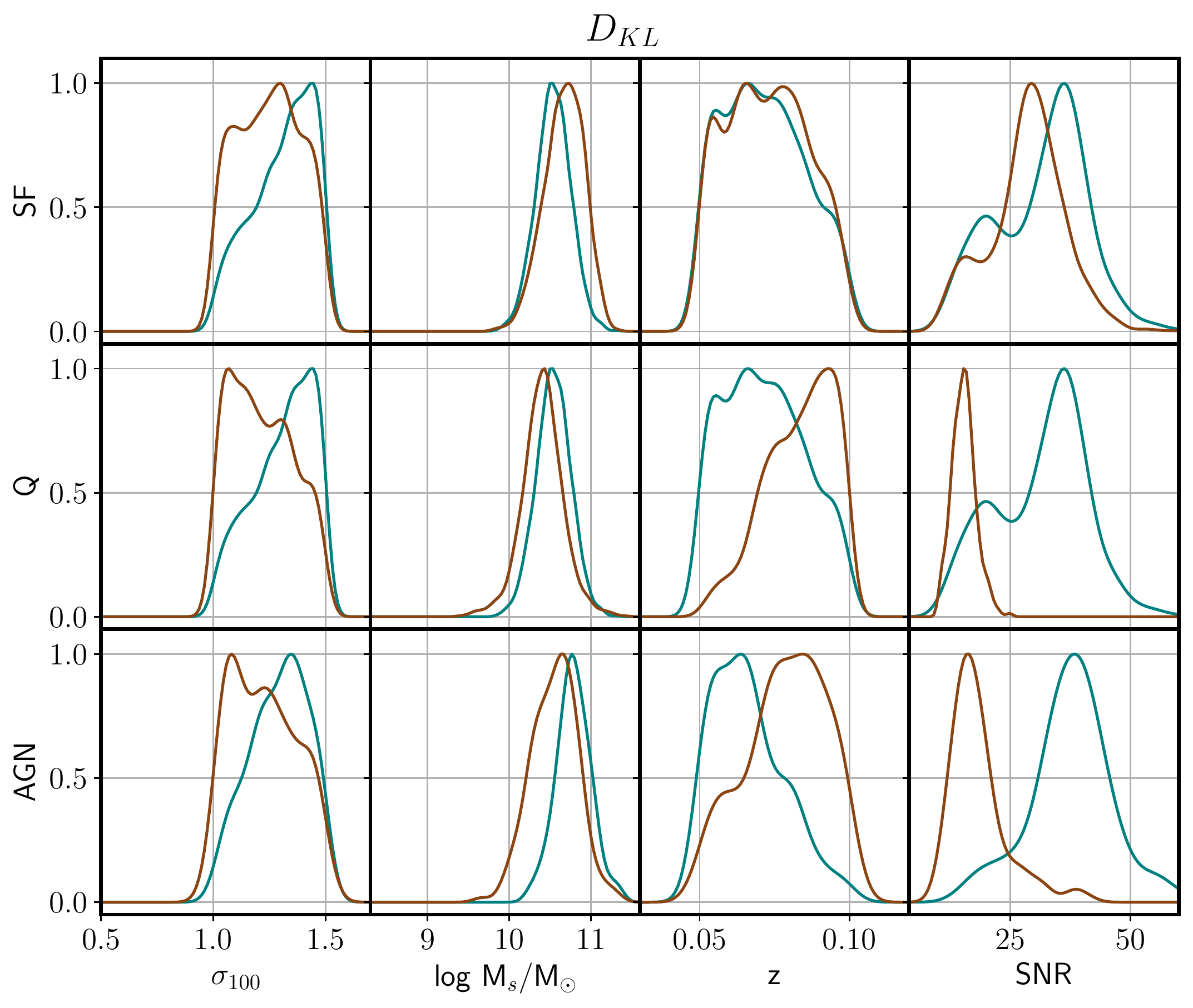}
  \caption{Distribution of some of the physical parameters
    of galaxies at the extreme ranges of PC1 (5th and 95th
    percentiles in teal blue and brown, respectively), according to
    negentropy ($\overline{\mathcal{H}}$, top) and
    relative entropy (D$_{\rm KL}$, bottom). From top to bottom
    in each figure, 
    the rows correspond to star-forming, quiescent and AGN spectra,
    respectively. From left to right, the measurements are 
    stellar velocity dispersion in units of 100\,km/s
    ($\sigma_{100}$), (log) stellar mass in solar units, redshift, and
    the average signal to noise ratio in the SDSS $r$ band.  The histograms
    are plotted with a Gaussian kernel density estimator.}
  \label{fig:histogHDkl}
\end{figure}

Fig.~\ref{fig:PCA_conts} shows the distribution of projections of the
observed spectra on the first two principal components for line strength data and
both definitions of
entropy, as labelled. The contours map the density from 50\% to 80\%
of total number of datapoints (in steps of 10\%), and are plotted
separately with respect to nebular activity, as above, into
star-forming (blue), quiescent (red), and AGN (green).  We emphasize
that the PCA output depends only on data in spectral regions where
emission line activity is absent or weak, so that most of the variance
cannot be directly ascribed to their classification into SF, Q and
AGN. The standard line strength data (LS, top) separate the subsamples along a diagonal
line in PC1-PC2 space, with AGN galaxies located between the SF and Q subsets.
Similarly, Shannon entropy (middle) shows substantial differences between the
SF and the Q subsets, with the AGN group populating an
intermediate area, something that hints at nuclear activity 
representing a transition from star formation to quiescence, as already 
proposed in the literature \citep[see, e.g.,][]{Schawinski:07, Salim:14}.
This diagram suggests that
the ``horizontal direction'', i.e.  PC1 (mainly dependent on D4K for $\overline{\mathcal{H}}$),
is the main discriminator, supporting the definition of the green valley,
i.e. transitioning galaxies from star formation to quiescence, using
this spectral feature \citep{JA:19, JA:20}.
In contrast, relative entropy (bottom) does 
not separate the spectra so well, and only produces an extended tail of
star-forming galaxies towards high, positive values of PC1.

Fig.~\ref{fig:histogHDkl}  shows the distribution of some
observables when taking subsets corresponding to the 5th (teal colour) and
95th (brown) percentiles of the distribution of PC1 projections
from negentropy ($\overline{\mathcal{H}}$) or relative
entropy (D$_{\rm KL}$), shown in both cases for
the star-forming (top), quiescent (middle) and AGN (bottom) samples.
We should stress that the parent sample is
restricted to a relatively narrow range of stellar velocity dispersion
(100-150\.km/s), so that the distributions are as ``homogeneous'' as
possible, concerning  variations derived from the general mass-age
and mass-metallicity trends \citep[see, e.g.,][]{Gallazzi:05}.
It is worth pointing out that this
analysis produces a noticeable segregation in the distribution of S/N -- a
property only dependent on the technical aspects of the observations --
in many instances, except for SF
galaxies. This is an important trend that all
methods based on deep / shallow learning should take into account. 
Taken at face value, it would imply that an important part of the
variance/information/signal one can extract from spectra could be due
to technical details of the data acquisition process.
Note that such segregation is not present 
in velocity dispersion, stellar mass, or redshift, for
$\overline{\mathcal{H}}$, although relative entropy also shows 
a redshift and velocity dispersion dependence. 
Appendix~\ref{App:SNR} performs the same analysis on a smaller
subset of galaxy spectra restricted at a higher S/N threshold,
showing that the overall trends are consistent.
This result illustrates the need to select clean, well-defined samples
in any statistical, data-driven study of the underlying star
formation history of galaxies. In addition, these methods can be applied
to separate out the observational effects from the physical ones -- 
as, in, e.g., the removal of night sky lines \citep{WH:05}.

\begin{figure*}
  {\Large
    \begin{equation}
      \qquad\qquad 
      \mathbb{C}_{\rm BC03}=
      \begin{blockarray}{ccccccc}
        D4K & H\delta & H\gamma & {\rm Mgb} & \langle{\rm Fe}\rangle & {\rm NaD} & \\[10pt]
        \begin{block}{(cccccc)c}
         1.     & -0.8705 & -0.8557 &  0.7263 &  0.7313 &  0.6483 & D4K\\[5pt]
        -0.8705 &  1.     &  0.9410 & -0.6839 & -0.6458 & -0.6114 & H\delta\\[5pt]
        -0.8557 &  0.9410 &  1.     & -0.6772 & -0.6303 & -0.6011 & H\gamma\\[5pt]
         0.7263 & -0.6839 & -0.6772 &  1.     &  0.5505 &  0.5012 & {\rm Mgb}\\[5pt]
         0.7313 & -0.6458 & -0.6303 &  0.5505 &  1.     &  0.5085 & \langle{\rm Fe}\rangle\\[5pt]
         0.6483 & -0.6114 & -0.6011 &  0.5012 &  0.5085 &  1.     & {\rm NaD}\\[5pt]
        \end{block}
      \end{blockarray}
      \label{eq:cov}
    \end{equation}
    \caption{Covariance matrix of the six line strengths targeted in the analysis of
      SDSS spectra, derived from the \citet{BC03} population synthesis models, renormalized
      to unit variance (i.e. defined as a correlation).
      The spectra have been convolved to an equivalent velocity dispersion
      of 100\,km\,s$^{-1}$, with Gaussian noise corresponding to S/N=10\,\AA$^{-1}$ in the SDSS-$r$ band,
      mimicking the observational data used in this paper. 
      Note the high degree of correlatedness between line strengths
      (see text for details).}
    \label{fig:cov}
  }
\end{figure*}

\section{Afterthought: The covariance of stellar population spectra}
\label{Sec:Cov}

The entropy-based analysis presented here illustrates the challenge of
extracting a unique solution to this particular inverse problem:
starting from a set of observed galaxy spectra, we aim at determining the
fundamental components, i.e. the base stellar populations that lead to
the underlying star formation histories.  While signal extraction
from superpositions is a
relatively easier task in, say, audio files or superposition of
independent time series, galaxy spectra reveal a high level of
entanglement regarding the information content as a function of
wavelength, that prevents us from extracting details of the age
distribution or chemical composition even in spectra with the highest
quality. In this epilogue, we show a simple exercise that makes use
of the optimal set of six spectral windows presented in this paper,
defined by the entropy content (see Fig.~\ref{fig:Hlam}).
Fig.~\ref{fig:cov} shows the covariance
matrix of a set of 16,386 synthetic spectra representing
simple stellar populations that cover a wide
range of age (0.5$<$t$_{\rm SSP}$$<$14\,Gyr) and metallicity 
($-$1.5$<$log\,Z/Z$_\odot$$<$$+$0.3),
from the models of \citet{BC03}.
Other sets of models produce very similar results. Each of the six indices is
renormalized to unit variance, to be able to compare all indices
in the same way. To produce data comparable to the SDSS spectra
used in this paper, we convolve the models to an equivalent velocity dispersion
of 100\,km\,s$^{-1}$, and add Gaussian noise corresponding to a
S/N in the SDSS-$r$ band of 10\,\AA$^{-1}$.
Note that all indices are correlated or anticorrelated
at a rather high level, amounting to $\gtrsim$70\% of the intrinsic variance,
being in some cases as high as $\sim$90\%.
Also note that these indices have been chosen to yield most of the
information content, or variance, across the optical window, and are
typically adopted in most analyses of stellar populations in the literature.
For instance, the 4000\AA\ break strength is often associated to an
overall stellar age, and Balmer absorption reveals episodes of star
formation within the past $\sim$1\,Gyr. However, the otherwise strong
metallicity sensitive indices Mgb, $\langle$Fe$\rangle$ and NaD
are also strongly correlated with the former, as expected from the
well-known age-metallicity degeneracy \citep[see, e,g,][]{Wo:94b,FCS:99}.

Observational errors and spectral resolution obviously affects the
outcome: the original models with no velocity dispersion of noise have an
even higher covariance.  This result should be taken as a note of caution
in the interpretation of any statistic that produces
a figure of merit based on the comparison of observations with
population synthesis models. The most commonly used one, $\chi^2$, is
expected to produce a minimum at the best fit, with a value roughly similar to
the number of degrees of freedom \citep{Andrae:10}. If we use the more standard definition of the
$\chi^2$ statistic for a generic covariance $\mathbb{C}$:
\begin{equation}
  \chi^2(\pi) \equiv \Big ( {\bf x} - {\bf m(\pi)}\Big )^T\cdot\mathbb{C}^{-1}\cdot
  \Big ({\bf x} - {\bf m(\pi)}\Big ),
\end{equation}
where ${\bf x}$ represents the measurements and ${\bf m}$ the model output
for a specific set of parameters ($\pi$). 
Perfectly uncorrelated data -- the assumption most often invoked -- 
implies a covariance $\mathbb{C}=\mathbb{I}$, and typically yields a
minimum $\chi^2$ around 6, whereas the covariance estimated from the model grid
produces a minimum $\chi^2$ of $\sim$20, clearly
inconsistent with the assumption that the six line strengths are independent
units of information. At face value, the substantially higher 
$\chi^2$ when accounting for covariance suggests that the fitting procedure 
effectively involves fewer degrees of freedom than the six (or more)
spectral features. Therefore, even though we may have a large number of
observable constraints from galaxy spectra, we can only 
produce a rough estimate
of stellar age, metallicity and possibly some non-solar abundance ratios.
This is not only applicable to line strength analysis,
but to full spectral fitting, where each 'pixel', i.e. a flux unit within
a relatively narrow spectral window ($\Delta\lambda$$\sim$1\,\AA), is highly
correlated with most of the
other pixels in the spectrum within the typical NUV-optical-NIR spectral
interval used in such analyses. Highly targeted studies of line strengths
are perhaps the only effective way to constrain the stellar population
content of galaxies, whereas methods based on full spectral fitting should
take into consideration this important caveat about covariance, and
should not make the assumption that the large number of flux measurements
are independent, a result that would falsely lead to a high constraining power.

\begin{figure*}
  {\Large
    \begin{equation}
      \qquad\qquad 
      \mathbb{C}_{\rm SDSS}=
      \begin{blockarray}{ccccccc}
        D4K & H\delta & H\gamma & {\rm Mgb} & \langle{\rm Fe}\rangle & {\rm NaD} & \\[10pt]
        \begin{block}{(cccccc)c}
        1.      & -0.5193 & -0.0420 &  0.7573 &  0.3901 &  0.3722 & D4K\\[5pt]
        -0.5193 &  1.     &  0.5249 & -0.3881 & -0.0042 & -0.1074 & H\delta\\[5pt]
        -0.0420 &  0.5249 &  1.     & -0.0290 &  0.1793 &  0.0551 & H\gamma\\[5pt]
        0.7573  & -0.3881 & -0.0290 &  1.     &  0.4445 &  0.3099 & {\rm Mgb}\\[5pt]
        0.3901  & -0.0042 &  0.1793 &  0.4445 &  1.     &  0.1841 & \langle{\rm Fe}\rangle\\[5pt]
        0.3722  & -0.1074 &  0.0551 &  0.3099 &  0.1841 &  1.     & {\rm NaD}\\[5pt]
        \end{block}
      \end{blockarray}
      \label{eq:cov}
    \end{equation}
    \caption{Equivalent covariance matrix as in Fig.~\ref{fig:cov},
    but derived from the observed SDSS spectra. Note the lower correlatedness of real galaxy spectra
    with respect to the models' covariance.}
    \label{fig:covSDSS}
  }
\end{figure*}

Equally interesting to the result obtained from the models
is a comparison with the covariance of the real
galaxy spectra from the SDSS sample (Fig.~\ref{fig:covSDSS}). While substantial,
the covariance is smaller than the one found in synthetic data.
This should come as no surprise. More realistic data regarding noise and the
``kinematic kernel'' will lower the covariance.
However, this  effect should be sub-dominant: For instance, the covariance
between D4k and H$\delta$ weakens from $-$0.9013 in the noiseless case to
$-$0.8705 at S/N=10, a change that is not large enough to explain the lower covariance of the
observations. A similar variation is obtained if we include the effect of
the stellar velocity dispersion. It is important to note that 
the synthetic models boil down to linear superpositions of $\sim$1,000 stellar
spectra (e.g. MILES, \citealt{MILES}). This result was, for instance, evident
when comparing the mutual information in a set of 2dFGRS spectra to data from
galaxy formation models \citep{IB:01}. A comparison of these two covariance matrices
should convince the reader that models still lack the complexity needed for
more detailed constraints of the star formation histories of galaxies.


\section{Conclusions}
\label{Sec:Conc}

This paper focuses on the fundamental limitations of extracting
information from galaxy spectra to derive star formation
histories. This work is complementary to the standard procedure 
based on comparisons of stellar population
synthesis models with the observational data via full spectral fitting
or targeted line strengths. There is a vast literature devoted
to the traditional methodology that perform careful analyses to
mitigate the inherent degeneracies among the population
properties 
\citep[to name a few:][]{MOPED:03,SL:05,STECMAP,VESPA:07,FSPS:10,ULySS:11,pPXF:12,FLB:13,Firefly}.
However, it is not the goal
of this paper to assess these methods, but to explore the 
limitations at a more fundamental level, based on the information
content (here defined as negentropy, alternatively defined as variance,
as in PCA-based work, e.g., \citealt{Rogers:07}).
The essence of the problem lies in the information content of the spectra,
that we base in this paper on entropy, either taking the fundamental
definition of \citet{SW:75} or relative entropy as defined by the
Kullback-Leibler divergence \citep{Dkl}. While alternative methods are
adopted in the literature to define information content, entropy
represents the 'building block' as it directly relates to, e.g. how
many bits of information are needed for a full representation of the
data \citep{Shannon:53}.  Entropy lies at the core of all methods
aimed at blind classification of galaxies, such as Principal Component
Analysis, Independent Component Analysis, Cross-entropy methods,
factor analysis, convolutional neural networks, etc.  We explore both
synthetic models and real galaxy spectra from SDSS, to find that a
reduced set of spectral regions encode most of the information,
unsurprisingly tracing the traditional Lick system
\citep[e.g.][]{Trager:98} commonly used in most studies.

The definition of entropy is initially determined for the full
spectral range of interest, say the optical window, and then narrower
intervals can be explored. Fig.~\ref{fig:Hlam} suggests that stellar
populations feature a reduced set of well-defined regions where the
information content is highest. Overall, the entropy of galaxy spectra
is rather high -- as in uninformative -- reflecting the difficult task
of constraining the details of the underlying populations, regardless
of the apparently high number of data points in each spectra.
Applying this method to SDSS galaxy spectra confirms this high entropy
(Fig.~\ref{fig:HlamSDSS}), with a clear, and fully expected,
dependence on nebular emission, with AGN systems representing an
intermediate stage between star forming and quiescent galaxies.  A
detailed analysis based on PCA applied to the entropy estimates
suggest that the 4000\AA\ break strength and Balmer absorption are the
most important sources of entropy variation, with metal-dependent
indicators being subdominant. This trend confirms previous work that
focuses on, e.g. the D$_n$(4000) vs H$\delta$ bivariate diagram as a
fundamental tool in the analysis of star formation
histories \citep[see, e.g.][]{Kauff:03}.  Intriguingly,
Fig.~\ref{fig:histogHDkl} shows that S/N may also affect the analysis
(even at S/N$\gtrsim$10\,\AA$^{-1}$), which implies that blind methods
are highly susceptible to the quality of the data.

Our results suggest that detailed estimates of star formation histories are 
hampered by the sizeable covariance of the spectral elements, which
result in an effectively low number of ``degrees of freedom'', regardless
of the large number of data units in a spectrum. Somehow, this addresses
the long-known fact that a reduced set of high quality
colours based on broadband photometry can produce constraints on the
stellar populations that are comparable with the analogous study at high
spectral resolution. For instance, the derivation
of stellar masses from photometry is comparable with the equivalent
analysis making use of spectra -- once the redshift is well
known \citep[e.g.][]{Santini:15}.
Moreover, non-solar abundance ratios produce variations 
over large spectral windows, especially at shorter wavelengths
\citep[e.g.][]{AV:15}, so that
broad- and medium-band photometry, in principle, carry such
detailed information. This would make surveys such as
J-PAS (\citealt{JPAS}, featuring 56 filters with 150\AA\ bandwidth),
or slitless grism spectroscopy (with e.g. ACS or WFC3
at the Hubble Space Telescope, NIRISS at the JWST or NISP at Euclid)
as informative, from the point of view of stellar populations,
as the more expensive spectral surveys. Studies of entropy variations
{\sl within} spectral data of the same galaxy -- from Integral Field Unit
observations -- can also be exploited to understand radial variations of
the underlying stellar populations, a concept that will be explored in
future work.


\section*{Acknowledgments}
IF acknowledges support from the Spanish Research Agency of the
Ministry of Science and Innovation (AEI-MICINN) under the grant with
reference PID2019-104788GB-I00. OL acknowledges STFC Consolidated
Grant ST/R000476/1 and a Visiting Fellowship at All Souls College,
Oxford.  This paper was meant to be produced during a visit to the
Flatiron Institute (FI), but it was not possible due to the SARS-CoV-2
pandemic. Nevertheless, IF warmly thanks the FI for offering to host his 
visit.  Funding for SDSS-III has been provided by the Alfred P. Sloan
Foundation, the Participating Institutions, the National Science
Foundation, and the U.S. Department of Energy Office of Science. The
SDSS-III web site is http://www.sdss3.org/.

\section*{Data availability}
This work has been fully based on publicly available data:
galaxy spectra were retrieved from the SDSS DR16 archive
(https://www.sdss.org/dr16/) and stellar population
synthesis models can be obtained from the respective authors.



\appendix

\section{Line strength covariance}
\label{App:LS}

\begin{figure}
  \centering
  \includegraphics[width=85mm]{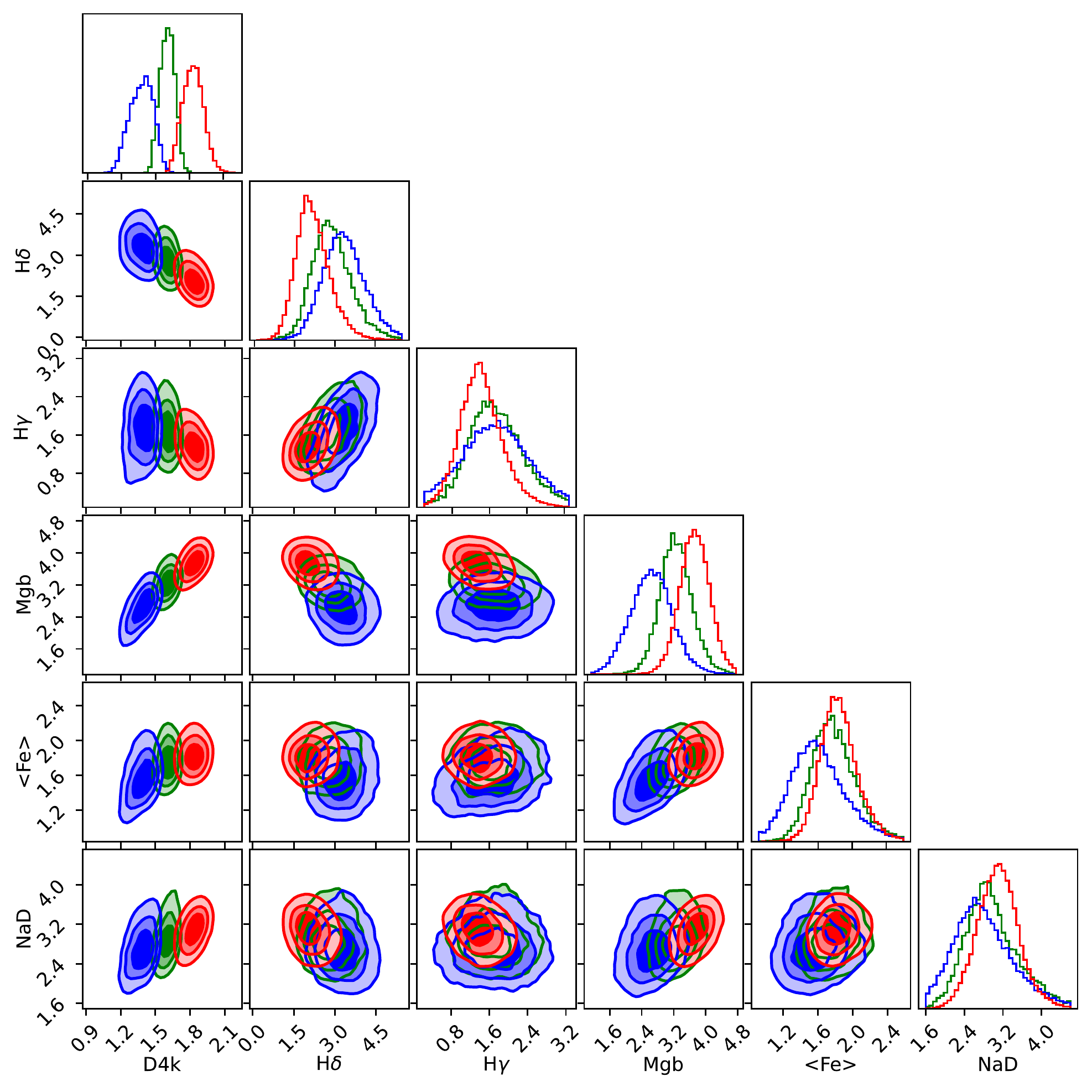}
  \caption{Distribution of observed line strengths in our
  working sample of SDSS galaxies. The contours engulf 25, 50 and 75\%
  of each subsample of star-forming (blue), AGN (green) and quiescent (red)
  galaxies.
  }
  \label{figA:LS}
\end{figure}

Fig.~\ref{figA:LS} shows a comparison of the distribution of line
strengths of galaxy spectra from SDSS in the six spectral regions
targeted in this paper. The data are colour-coded according to
nebular emission into star-forming (blue), AGN (green) and quiescent
(red), see main text for details. Note the strong correlation between the
4000\AA\ break strength and either H$\delta$ or Mgb, as can also
be seen in the covariance matrix of the SDSS spectra (Fig.~\ref{fig:covSDSS}).

\section{Selecting a sample at higher S/N}
\label{App:SNR}

\begin{figure}
  \centering
  \includegraphics[width=85mm]{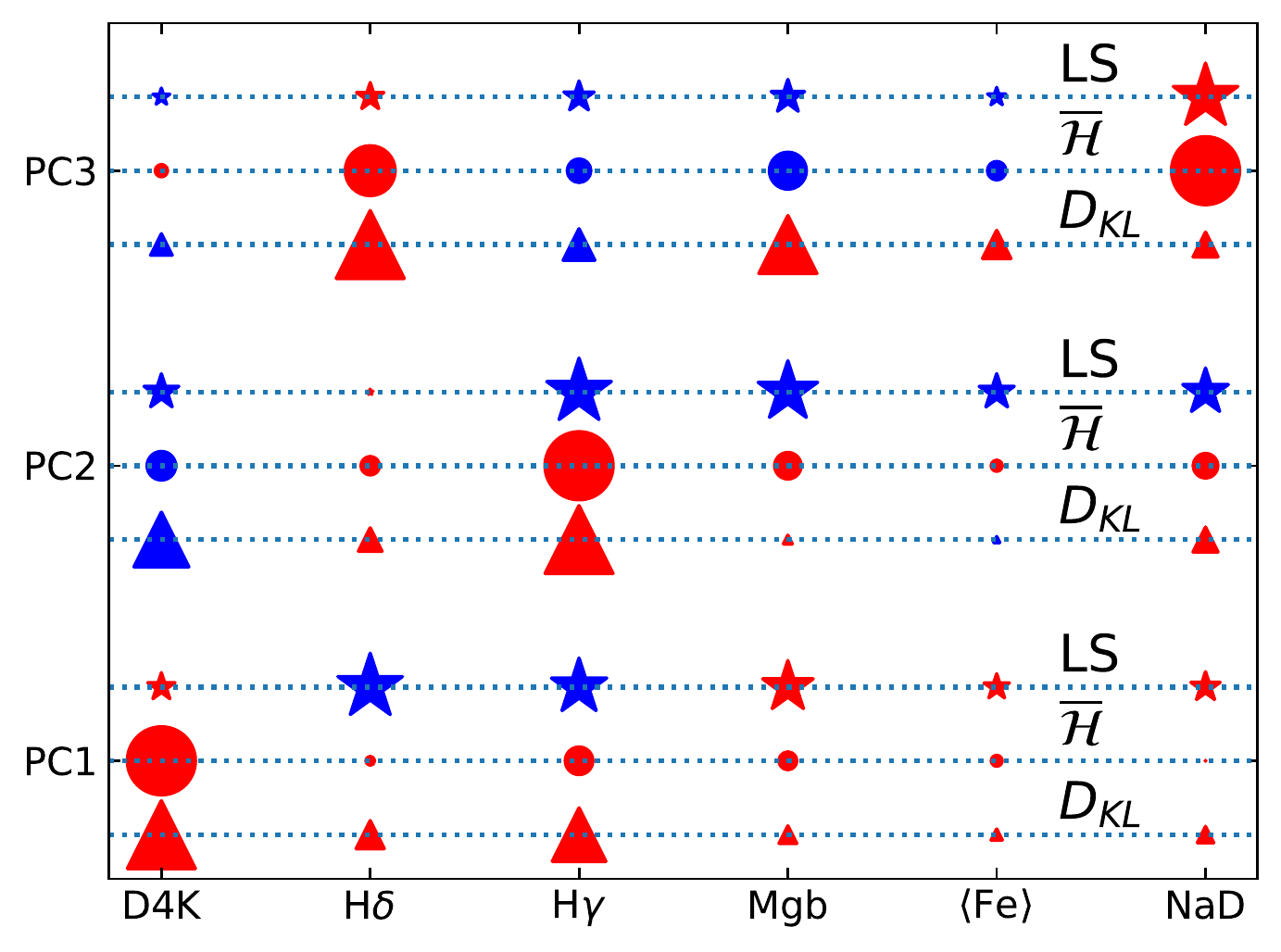}
  \caption{Equivalent of Fig.~\ref{fig:PCA_weight}
  for a subsample of spectra restricted to S/N$>$30 (instead of 10 in
  the original sample).}
  \label{figA:PCA_weight}
\end{figure}

\begin{figure}
  \centering
  \includegraphics[width=85mm]{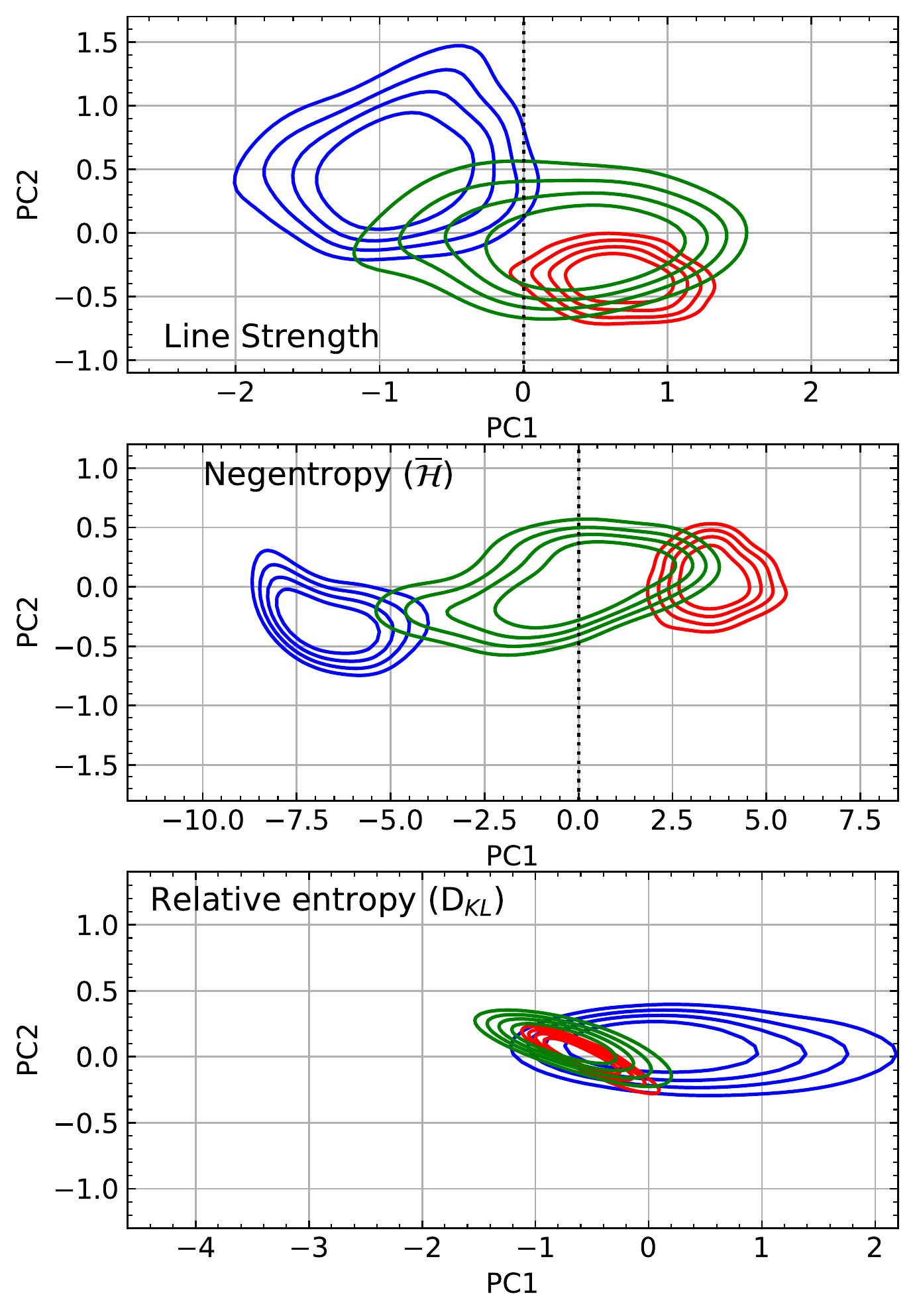}
  \caption{Equivalent of Fig.~\ref{fig:PCA_conts} for a
  subsample of spectra restricted to S/N$>$30 (instead of 10 in
  the original sample).}
  \label{figA:PCA_conts}
\end{figure}

\begin{figure}
  \centering
  \includegraphics[width=85mm]{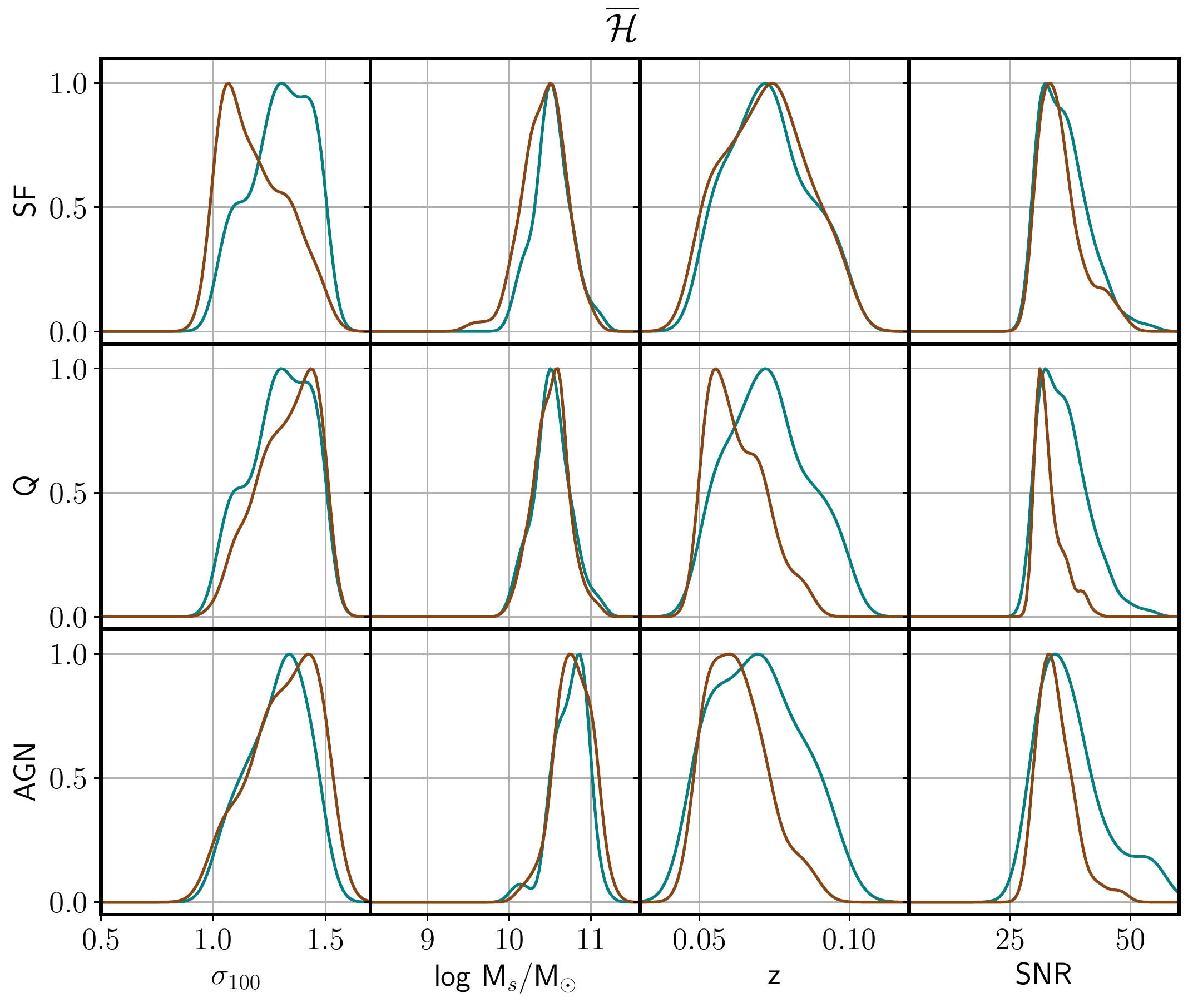}
  \includegraphics[width=85mm]{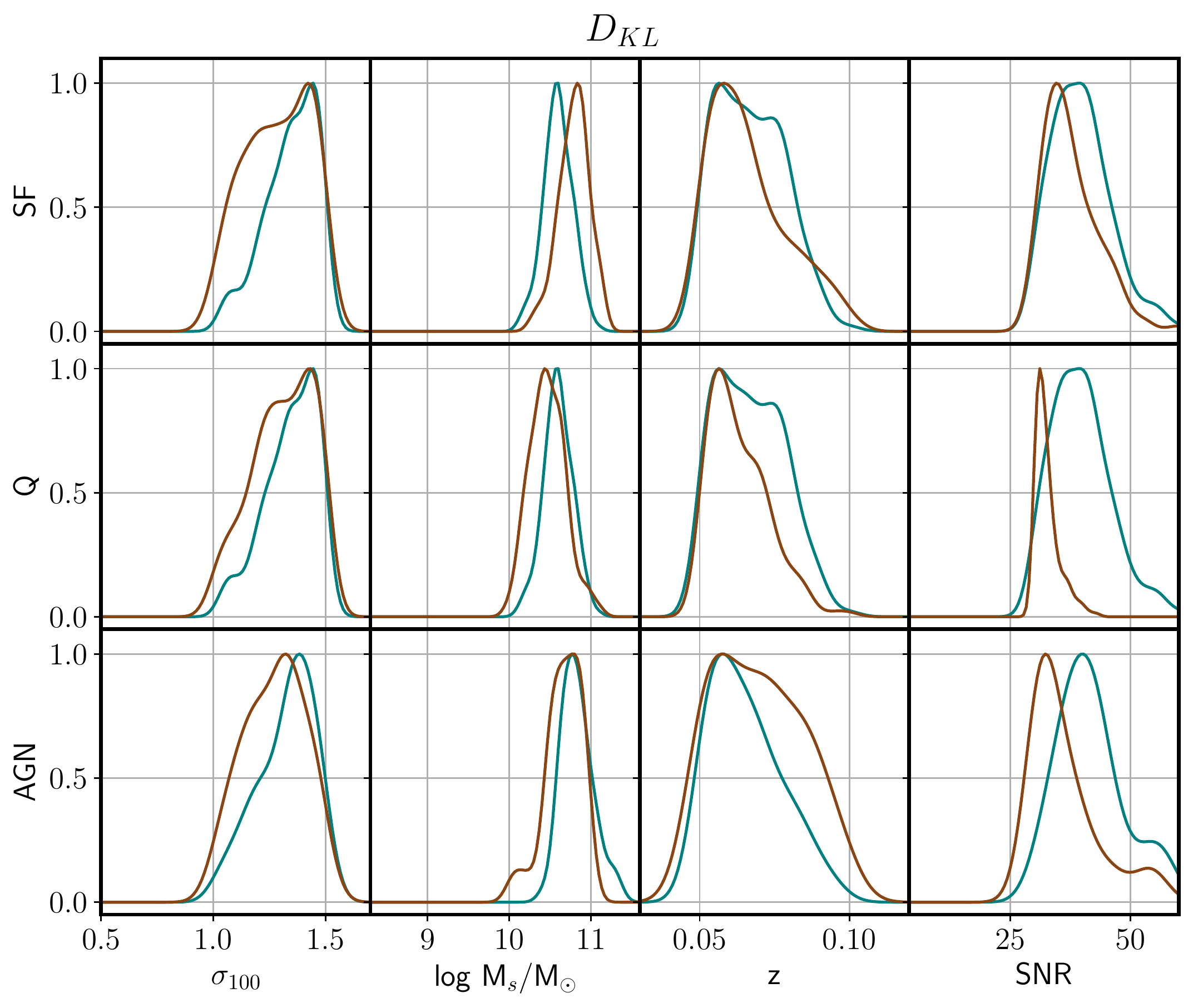}
  \caption{Equivalent of Fig.~\ref{fig:histogHDkl} for a
  subsample of spectra restricted to S/N$>$30 (instead of 10 in
  the original sample).}
  \label{figA:histogHDkl}
\end{figure}

In Fig.~\ref{fig:histogHDkl} we saw that the distribution of galaxy
spectra at extreme ends of the projection onto PC1 feature a
significant correlation with S/N. While this could be an indirect trend
caused by the sample selection, we reanalyse here a subset of spectra
with a more stringent cut on S/N ($>$30, in contrast with S/N$>$10 for
the original sample). The sample decreases by a factor $\sim$7, from
the original 76,569 spectra to 10,574, still a reasonable number for
our statistical analysis. The results are presented in
Figs.~\ref{figA:PCA_weight}, \ref{figA:PCA_conts},
and \ref{figA:histogHDkl}.  The eigenvalues of the PCA in this subset
are given in Table~\ref{tabA:scree}, and the principal components are
quantified in Table~\ref{tabA:weights}. The results are very similar
to those presented in the main text for the original sample.
The eigenvalues show that the variance is, once more,
dominated by $\sim$2 principal
components when using entropy estimates, and appears more extended
with the traditional line strengths -- requiring, for instance,
four components with the line strength inputs to engulf 98\% of variance,
whereas entropy (relative entropy) only require two (three) components. This
subset also depends on S/N in a similar way, although the sample
appears culled at low velocity dispersion -- understandably from the
fact that lower $\sigma$ galaxies are overall fainter and thus will
correspond to lower S/N in the SDSS spectra. It is also worth noting
the redshift trend in Q and AGN spectra, as the higher S/N produces an
effectively shallower survey, and is thus more prone to Malmquist bias.
Consistently to the original sample, the principal components also
feature the 4000\AA\ break strength as the dominant one (PC1), but the second
component is now mostly dependent on H$\gamma$ (for both definitions
of entropy). The contribution from H$\delta$ is now relegated to PC3, in
combination with NaD. In any case, the distribution of principal
component projections (Fig.~\ref{figA:PCA_conts}) is similar to the
original case, suggesting that while S/N should be taken into account,
it does not dominate the variations found in the data.
 
\begin{table}
\caption{Normalized PCA eigenvalues (high S/N)\label{tabA:scree}}
\begin{center}
\begin{tabular}{cccc}
\hline
Rank & \multicolumn{3}{c}{Eigenvalue ratio}\\
 & Line Strength & Entropy & D$_{\rm KL}$\\
\hline
1 & 0.52706 & 0.97671 & 0.89443\\
2 & 0.29045 & 0.01364 & 0.07126\\
3 & 0.12832 & 0.00379 & 0.01337\\
4 & 0.03983 & 0.00278 & 0.01067\\
5 & 0.01074 & 0.00223 & 0.00556\\
6 & 0.00360 & 0.00085 & 0.00472\\
\hline
\end{tabular}
\end{center}
\end{table}

\begin{table*}
\caption{The first three PCs of the high S/N subsample (normalized)\label{tabA:weights}}
\begin{center}
\begin{tabular}{l@{\quad\vline\quad}ccc@{\quad\vline\quad}ccc@{\quad\vline\quad}ccc}
\hline
 & \multicolumn{3}{c@{\vline\quad}}{LS} & \multicolumn{3}{c@{\vline\quad}}{$\overline{\mathcal{H}}$}& \multicolumn{3}{c}{D$_{\rm KL}$}\\
\hline
Region & PC1 & PC2 & PC3 & PC1 & PC2 & PC3 & PC1 & PC2 & PC3\\
\hline
D4K &  0.12786 & $-$0.19873 & $-$0.06482 & 0.98395 & $-$0.17063 &  0.02782 &  0.83637 & $-$0.54142 & $-$0.07980\\
H$\delta$ & $-$0.70398 &  0.00536 &  0.16915 & 0.01693 &  0.07190 &  0.45532 &  0.14586 &  0.10089 &  0.77944\\
H$\gamma$ & $-$0.51732 & $-$0.68209 & $-$0.20733 & 0.16201 &  0.96280 & $-$0.10210 &  0.52213 &  0.82680 & $-$0.17114\\
Mgb &  0.43158 & $-$0.58063 & $-$0.25118 & 0.06814 &  0.14675 & $-$0.25022 &  0.05906 &  0.01503 &  0.57133\\
$\langle$Fe$\rangle$ &  0.11080 & $-$0.20273 & $-$0.07546 & 0.02570 &  0.02606 & $-$0.06314 &  0.02399 & $-$0.00719 &  0.13797\\
NaD &  0.14799 & $-$0.34204 &  0.92488 & 0.00016 &  0.12854 &  0.84551 &  0.05027 &  0.11318 &  0.10659\\
\hline
\end{tabular}
\end{center}
\end{table*}


\bsp	
\label{lastpage}
\end{document}